\documentclass[aps,pre,preprint, showpacs, superscriptaddress]{revtex4}

\usepackage[dvips]{graphicx}
\usepackage{latexsym}
\usepackage{natbib}

\begin{document}

\newcommand{\oiint}[1]{\int \hspace{-0.25cm} \int_{#1}
\hspace{-0.61cm} \bigcirc \; }

\bibliographystyle{apsrev}

\title{The Taylor Interpolation through FFT Algorithm for   Electromagnetic Wave Propagation and Scattering  }

\author{Shaolin  Liao}
  \affiliation{Department of Electric and Computer
Engineering, University of Wisconsin, Madison, \\ 1415 Engineering Drive, Madison, WI, USA, 53706  \\
E-mail: sliao@wisc.edu }

\today

\begin{abstract}
The Taylor Interpolation through  FFT (TI-FFT) algorithm for the
computation of the electromagnetic wave propagation in the
quasi-planar geometry
 within the half-space is proposed in this article.  There are two types of  TI-FFT algorithm, i.e.,
 the {spatial} TI-FFT  and
  the {spectral} TI-FFT. The former works in the spatial domain and
  the latter works in the spectral domain. It has been shown that
  the optimized computational complexity is the same for both types
  of  TI-FFT algorithm, which is $\mathcal{N}_r^{\hbox{\tiny opt}}
   \mathcal{N}_o^{\hbox{\tiny opt}}  \mathcal{O} (N \log_2
N)$ for an $N = \mathcal{N}_x \times \mathcal{N}_y$ computational
grid, where $\mathcal{N}_r^{\hbox{\tiny opt}}$ is the optimized
number of slicing reference planes  and $\mathcal{N}_o^{\hbox{\tiny
opt}}$ is the optimized order of Taylor series. Detailed analysis
shows that $\mathcal{N}_o^{\hbox{\tiny opt}}$ is closely related to
the algorithm's computational accuracy $\gamma_{\hbox{\tiny TI}}$,
which is given as $\mathcal{N}_o^{\hbox{\tiny opt}} \sim - \ln
\gamma_{\hbox{\tiny TI}}$ and the optimized spatial slicing spacing
between two adjacent spatial reference planes $\delta_z^{\hbox{\tiny
opt}}$  only depends on the characteristic wavelength  $\lambda_c$
of the electromagnetic wave, which is given as
$\delta_z^{\hbox{\tiny opt}} \sim \frac{1}{17} \lambda_c$. The
planar TI-FFT algorithm allows  a large sampling spacing required by
the sampling theorem. What's more,  the algorithm is free of
singularities and it works particularly well for the narrow-band
beam and the quasi-planar geometry.
\end{abstract}



\maketitle

\section{Introduction}
\label{sec:introduction}

 The computation  of electromagnetic wave propagation using the direct integration method is not efficient for the
large-scale computation because the direct integration method has a
daunting computational complexity of $\mathcal{O} \left(N^2\right)$
for an $N = \mathcal{N}_x \times \mathcal{N}_y$ computational grid,
e.g., in the beam-shaping mirror system design for the Quasi-Optical
(QO) gyrotron application, days of computation is required
\cite{Shaolin_conf,Rong,Perkins,Shaolin_JEMWA,Shaolin_ISAPE}.
Fortunately, when the computational geometry is a plane, the FFT has
been shown to be efficient in the electromagnetic wave computation
\cite{Tukey,Oppenheim,Wang}, which has  a computational complexity
of ${\mathcal{O}(N \log_2 N)}$ and a low sampling rate only limited
by the Nyquist rate.  For   the quasi-planar geometry, it will be
shown in this article that the FFT can still be used with the help
of the Taylor Interpolation (TI) technique.

\begin{figure}[h]\centering
\hspace{-0.2in}\includegraphics[width=4.7 in, height= 3.3
in]{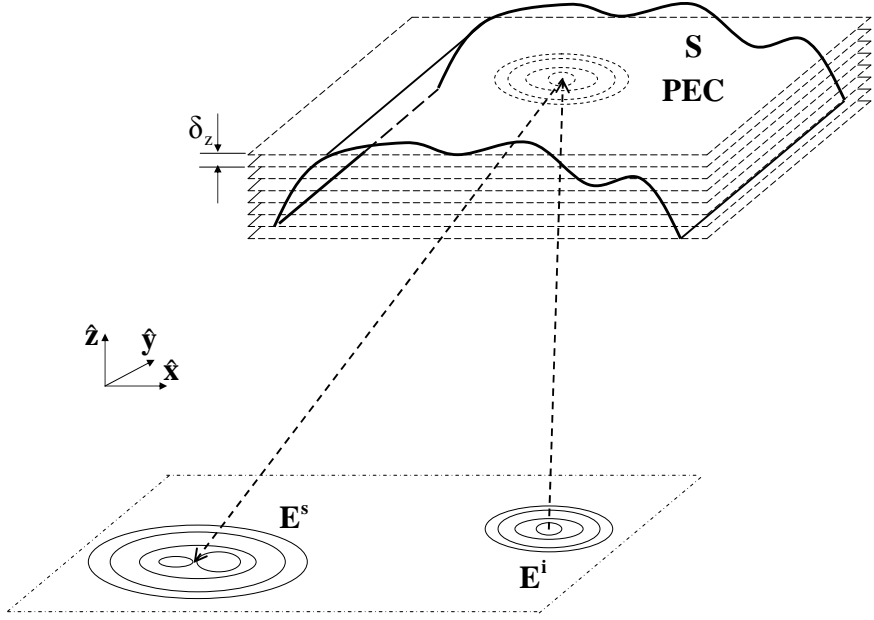} \caption{ Electromagnetic wave propagation and
scattering: the computation of the electromagnetic wave propagation
(the incident file ${\bf E}^i$) onto  the PEC surface $S$ is
implemented through the spatial TI-FFT and the computation of  the
scattered electromagnetic field ${\bf E}^s$ from the PEC surface $S$
is implemented through the   spectral TI-FFTs and the inverse
Fourier transform. $\delta z$ is the spatial slicing spacing in the
spatial TI-FFT. }
 \label{fig:scheme}
\end{figure}

 The rest of this article is organized as follows.
Section \ref{sec:spectral} gives  the 2-Dimensional (2D) Fourier
spectrum of the electromagnetic wave  in its closed-form expression.
Section \ref{sec:TIFFT} presents the optimized {spatial} and
{spectral} types of TI-FFT algorithm. In Section \ref{results}, one
numerical example is used to show the performance of the planar
TI-FFT algorithm. Section \ref{sec:discussion} discusses the
advantages and problems of the planar TI-FFT algorithm; some helpful
suggestions are given. Finally, Section \ref{sec:conclusion}
summarizes the planar TI-FFT algorithm. The scheme used to
illustrate the planar TI-FFT algorithm is shown in Fig.
\ref{fig:scheme} and the time dependence $e^{j \omega t}$ has been
assumed in this article.

\section{Electromagnetic Wave in the Spectral
Domain}\label{sec:spectral}

In this section, the 2D Fourier spectrum and far-field of the
electromagnetic wave for the radiation integral are shown to be
closely related to each other.

\subsection{The radiation integral}\label{subsec:radiation}
For given electric and magnetic surface currents   (${\bf J}_s, {\bf
J}_{ms}$), the radiating electric field ${\bf E}$ can be obtained
under the Lorenz condition \cite{Collin,Balanis}, which is given as

 \vspace{-0.1 in}

\begin{equation}\label{radiation}
{\bf E} =  \frac{-j}{\omega  \epsilon} \int \!\!\!
\int_S \left[ \hspace{-0.05in} \begin{array}{cccc}  \\ \\
 \end{array} k^2 {\bf
J}_s({\bf r}') G({\bf R}) + \left(\hspace{-0.05in}
\begin{array}{cccc}  \\ \\
 \end{array} {\bf J}_s({\bf r}') \cdot
\nabla' \hspace{-0.05in} \begin{array}{cccc}  \\
 \end{array}\right) \nabla' G({\bf R}) \hspace{-0.05in}
 \begin{array}{cccc}
 \\
 \end{array}  - j \omega \epsilon {\bf J}_{ms}({\bf r}') \times \nabla'
G({\bf R}) \hspace{-0.05in} \begin{array}{cccc}  \\  \\
 \end{array} \right] dV',
 \end{equation}

\hspace{-0.22in}where, $\nabla'$ is the gradient operator on the
source coordinate ${\bf r}'$ and  the scalar Green's function is
given as

 \vspace{-0.2in}

\begin{eqnarray}\label{scalargreen}
G({\bf R}) =  \frac{e^{-j k | {\bf R} |}}{4 \pi |{\bf R}|},
\hspace{0.4in}  {\bf R} \equiv {\bf r} - {\bf r}'.
\end{eqnarray}

\subsection{The 2D Fourier spectrum of the scalar Green's function}\label{subsec:green}

Now apply  the 2D Fourier transform on the scalar Green's function
$G({\bf R})$ in (\ref{scalargreen}),

    \vspace{-0.1in}

\begin{eqnarray}\label{Green2DFT}
  {\mathcal{G}}(k_x,k_y,{\bf r}')  \equiv \hbox{\large FT}_{\hbox{\tiny 2D}}  \left[ \hspace{-0.05in} \begin{array}{cc}  \\
 \end{array}    G({\bf R})   \hspace{-0.05in} \begin{array}{cccc}
\\  \\
 \end{array} \right]  =   \frac{1}{2\pi} \int_{x=-\infty}^\infty
 \int_{y=-\infty}^\infty  \frac{e^{-j k |{\bf R}|}}{4 \pi |{\bf R}|}
 e^{j k_x x}  e^{j k_y y}
  dx dy,
  \end{eqnarray}

\hspace{-0.22in}where $k_z$ and the 2D Fourier transform has been
defined as

  \vspace{-0.1 in}

\begin{eqnarray}\label{kz}
    k_z = \left\{  \begin{array}{cc}   \sqrt{k^2 - k_x^2 - k_y^2}, \hspace{0.6in} k_x^2 + k_y^2 < k^2    \\
    -j\sqrt{ k_x^2 + k_y^2 -k^2}, \hspace{0.43in} k_x^2 + k_y^2 \ge k^2   \end{array}
    \right.,
\end{eqnarray}

  \vspace{-0.1in}

\begin{eqnarray}\label{2DFTdef}
  \hbox{\large FT}_{\hbox{\tiny 2D}}  \left[\hspace{-0.05in} \begin{array}{cccc}  \\
 \end{array} \cdot  \hspace{-0.05in} \begin{array}{cccc}  \\
 \end{array}
 \right]  =  \frac{1}{2 \pi} \int_{x=-\infty}^\infty  \left\{    e^{j k_x
 x}
\int_{y=-\infty}^\infty
\left[\hspace{-0.05in} \begin{array}{cccc}  \\
 \end{array}  \cdot    \hspace{-0.05in} \begin{array}{cccc}  \\
 \end{array} \right]    e^{j k_y y} dy   \right\} dx,
\end{eqnarray}

From (\ref{Green2DFT}),

 \vspace{-0.1in}

\begin{eqnarray}\label{Green2DFT1}
  {\mathcal{G}}(k_x,k_y,{\bf r}')  = \frac{1}{2\pi}  e^{j k_x x'}  e^{j k_y y'} \int_{x=-\infty}^\infty
   \left\{  \hspace{-0.05in} \begin{array}{cccc}
 \\  \\
 \end{array}   e^{j k_x (x-x')}    \right.    \times  \int_{y=-\infty}^\infty  \left[ \frac{e^{-j k |{\bf R}|}}{4 \pi |{\bf R}|}  \left.
     e^{j k_y (y-y')}
   \hspace{-0.05in} \begin{array}{cccc}  \\  \\
 \end{array}  \right]   dy \right\} dx,
\end{eqnarray}

Changing variables $u=x-x'$, $v=y-y'$ and  $w=z-z'$,
(\ref{Green2DFT1}) becomes,

 \vspace{-0.1in}

\begin{eqnarray}\label{der}
 \hbox{\large FT}_{\hbox{\tiny 2D}}  \left[ \hspace{-0.05in} \begin{array}{cc}
\\ \\
 \end{array}    G({\bf R})   \hspace{-0.05in} \begin{array}{cccc}  \\
 \end{array} \right]
  =   \frac{1}{2\pi}  e^{j k_x x'}  e^{j k_y y'}
  \int_{u=-\infty}^\infty  \left\{  \hspace{-0.05in} \begin{array}{cccc}
 \\  \\
 \end{array}  e^{j k_x u}  \times  \int_{v=-\infty}^\infty  \left[
\hspace{-0.05in} \begin{array}{cccc}
 \\  \\
 \end{array} \frac{e^{-j k |{\bf R}|}}{4 \pi |{\bf R}|}
  e^{j k_v v}
    \right]  dv  \right\} du
\end{eqnarray}

In the cylindrical coordinate,

 \vspace{-0.1in}

\begin{eqnarray}\label{der0}
|{\bf R}|
 =\sqrt{(r_{\perp})^2 + w^2}
\end{eqnarray}

\hspace{-0.22in}where $r_{\perp} = u^2 + v^2$ and the following
relation can be obtained from (\ref{der0}),

 \vspace{-0.1in}

\begin{eqnarray}\label{der1}
 d  r_{\perp}   = \frac{|{\bf R}|}{r_{\perp}} d |{\bf
 R}|
\end{eqnarray}

Now, express (\ref{der}) in the cylindrical coordinate with the help
of (\ref{der1}),

 \vspace{-0.1in}

\begin{eqnarray}\label{der2}
 {\mathcal{G}}(k_x,k_y,{\bf r}')
 =  \frac{1}{4\pi}   e^{j k_x x'}  e^{j k_y y'}   \int_{|{\bf
R}|=|w|}^\infty      \left\{  \hspace{-0.05in} \begin{array}{cc}
\\  \\
 \end{array}   e^{-j k |{\bf R}|}   \frac{1}{2 \pi }  \int_{\phi=0}^{2 \pi}  \left[
  \hspace{-0.05in} \begin{array}{cc}     \\  \\
 \end{array}
    \ {e^{-j k_{\perp} r_{\perp} \cos (\psi-\phi)
 }}   \right]    d \phi  \right\}  d |{\bf  R}|
\end{eqnarray}

\hspace{-0.22in}where $\psi = \arctan\left[ \frac{k_y}{k_x} \right]$
and $\phi = \arctan\left[ \frac{v}{u} \right]$. The integration over
$\phi$ is  the Bessel function of the first kind of order 0 and
(\ref{der2}) reduces to

\vspace{-0.1in}

\begin{eqnarray}\label{der3}
 {\mathcal{G}}(k_x,k_y,{\bf r}')
 & = &  \frac{1}{4\pi} e^{j k_x x'}  e^{j k_y y'}   \int_{|{\bf
R}|=|w|}^\infty \left[  \hspace{-0.05in} \begin{array}{cc}     \\
\\
 \end{array}   e^{-j k |{\bf R}|}    \times J_0\left( \hspace{-0.05in} \begin{array}{cc}     \\
 \end{array}  k_{\perp } \sqrt{|{\bf R}|^2 - w^2 } \hspace{-0.05in} \begin{array}{cc}
 \\  \\
 \end{array} \right) \right]  d |{\bf
 R}|  \nonumber \\
 & = &  \frac{-j }{ 4 \pi k_z}  e^{j k_x x'}  e^{j k_y y'}  e^{-j k_z
 |z-z'|}
\end{eqnarray}

Because only half-space $z > z'$ is of interest, only the 2D Fourier
spectrum for half-space $z > z'$ will be considered in the rest of
this article, which is obtained from (\ref{der3}) as

\vspace{-0.1in}

\begin{eqnarray}\label{2DFT>}
 {\mathcal{G}}^>(k_x,k_y,{\bf r}')
   =  \frac{-j }{ 4 \pi k_z}  e^{j {\bf k} \cdot  {\bf r}'}  e^{-j k_z
z}
 \end{eqnarray}

\subsection{2D Fourier spectra of  Green's function related expressions}\label{subsec:epxr}

 The 2D Fourier spectra of the derivatives (order $n$) of
the scalar Green's function  can be obtained from the property of
the Fourier transform \cite{Oppenheim},

\vspace{-0.1in}

\begin{eqnarray}\label{spectrumdxyn}
 \frac{\partial ^{(n)} G({\bf R})}{\partial \tau^{(n)}}  \Longrightarrow  (-j k_\tau)^{n} {\mathcal{G}^>}(k_x,k_y,{\bf r}'),
\hspace{0.2in} \tau=x,y,z.
\end{eqnarray}

 Particularly, for the first-order and second-order derivatives,

\vspace{-0.1in}

\begin{eqnarray}\label{spectrumdxy}
  \frac{\partial  G({\bf R})}{\partial \tau }   \Longrightarrow
  \frac{-k_\tau}{4 \pi k_z}
e^{j {\bf k} \cdot  {\bf r}'}  e^{-j k_z z}, \hspace{0.1in}
\tau=x,y,z.
\end{eqnarray}

\vspace{-0.1in}

\begin{eqnarray}\label{spectrumdxy2}
  \frac{\partial^2  G({\bf R})}{\partial \tau^2 }   \Longrightarrow
    \frac{j k_\tau^2}{4 \pi k_z}
e^{j {\bf k} \cdot  {\bf r}'}  e^{-j k_z z}, \hspace{0.1in}
\tau=x,y,z.
\end{eqnarray}

Similarly, the 2D Fourier spectra of the following expressions can
be obtained for half-space $z > z'$,

\vspace{-0.1in}

\begin{eqnarray}\label{gradient}
 \hbox{\large FT}_{\hbox{\tiny 2D}}  \left[\hspace{-0.05in} \begin{array}{cccc}  \\
 \end{array}  \nabla   G({\bf R})  \hspace{-0.05in} \begin{array}{cccc}  \\
 \end{array}
 \right] \Longrightarrow  - j {\bf k}  {\mathcal{G}^>}(k_x,k_y,{\bf r}').
\end{eqnarray}

\vspace{-0.1in}

\begin{eqnarray}\label{laplace}
 \hbox{\large FT}_{\hbox{\tiny 2D}}  \left[\hspace{-0.05in} \begin{array}{cccc}  \\
 \end{array}  \nabla^2   G({\bf R})  \hspace{-0.05in} \begin{array}{cccc}  \\
 \end{array}
 \right]  \Longrightarrow  - k^2  {\mathcal{G}^>}(k_x,k_y,{\bf r}').
\end{eqnarray}

\vspace{-0.1in}

\begin{eqnarray}\label{tensor}
 \hbox{\large FT}_{\hbox{\tiny 2D}}  \left[\hspace{-0.05in} \begin{array}{cccc}  \\
 \end{array}  \nabla \nabla    G({\bf R})  \hspace{-0.05in} \begin{array}{cccc}  \\
 \end{array}
 \right]  \Longrightarrow  - {\bf k} {\bf k} {\mathcal{G}^>}(k_x,k_y,{\bf r}').
\end{eqnarray}

\vspace{-0.1in}

\begin{eqnarray}\label{dyadice}
 \hbox{\large FT}_{\hbox{\tiny 2D}}  \left[\hspace{-0.05in} \begin{array}{cccc}  \\
 \end{array}      \overline{\overline{G}}_e({\bf R})  \hspace{-0.05in} \begin{array}{cccc}  \\
 \end{array}
 \right]  \Longrightarrow   {\mathcal{G}^>}(k_x,k_y,{\bf r}') \left[\overline{\overline{I}} - \frac{{\bf k} {\bf k}}{k^2}
 \right].
\end{eqnarray}

\vspace{-0.1in}

\begin{eqnarray}\label{dyadich}
 \hbox{\large FT}_{\hbox{\tiny 2D}}  \left[\hspace{-0.05in} \begin{array}{cccc}  \\
 \end{array}      \overline{\overline{G}}_m({\bf R})  \hspace{-0.05in} \begin{array}{cccc}  \\
 \end{array}
 \right]  \Longrightarrow  -j  {\mathcal{G}^>}(k_x,k_y,{\bf r}') \left[  {\bf k} \times \overline{\overline{I}}
 \right].
\end{eqnarray}

\hspace{-0.22in}where the dyadic Green's functions of the electric
type ($\overline{\overline{G}}_e$) and the magnetic type
($\overline{\overline{G}}_m$) are given as

\vspace{-0.1in}

\begin{eqnarray}\label{dyadicE}
 \overline{\overline{G}}_e({\bf R}) =
 \left( \overline{\overline{I}} + \frac{1}{k^2} \nabla \nabla
 \right) G ({\bf R}).
\end{eqnarray}

\vspace{-0.1in}

\begin{eqnarray}\label{dyadicH}
 \overline{\overline{G}}_m({\bf R}) =  \nabla
  G ({\bf R})  \times \overline{\overline{I}}.
\end{eqnarray}

\subsection{The  far-fields}\label{FF}

In the far-field limit, ${\bf R} \simeq {\bf r}  \rightarrow
\infty$,

\vspace{-0.1in}

\begin{eqnarray}\label{FFGreen}
 G({\bf r})   =   \frac{e^{-j k | {\bf r} |}}{4 \pi |{{\bf r}|}},
 \hspace{0.4in} {\bf R}= {\bf r} \ \hbox{in the far-field limit}.
\end{eqnarray}

Similarly, the first-order derivative of the Green's function in the
far-field limit can be obtained as

\vspace{-0.1in}
\begin{eqnarray}\label{FFGreendxyz}
 \frac{\partial G({\bf r})}{\partial \tau } = \frac{\tau}{|{{\bf r}|}}
\left(  -j k -\frac{1}{|{{\bf r}|}^2} \right) \frac{e^{-j k | {\bf
r} |}}{4 \pi |{{\bf r}|}}  \simeq   - j \left(  \frac{\tau}{|{{\bf
r}|}}  k \right) \frac{e^{-j k | {\bf r} |}}{4 \pi |{{\bf r}|}} = -
j k_\tau \frac{e^{-j k | {\bf r} |}}{4 \pi |{{\bf r}|}},
\hspace{0.in} \tau = x, y, z
\end{eqnarray}

\hspace{-0.22in}where only $\frac{1}{|{\bf r}|}$ term is kept and
the other terms ($\frac{1}{|{\bf r}|^2}, \frac{1}{|{\bf r}|^3}$,
$\cdot \cdot \cdot$) are ignored. In derivation of
(\ref{FFGreendxyz}), the following relation has been used in the
far-field limit,

\vspace{-0.1in}

\begin{eqnarray}\label{FFlimit}
\frac{k_\tau}{k} = \frac{\tau}{ |{{\bf r}}|}, \hspace{0.5in} \tau =
x, y, z.
\end{eqnarray}

Following the similar procedure given in (\ref{FFGreendxyz}), the
far-fields of derivatives (order $n$) of the scalar Green's function
are obtained as

\vspace{-0.1in}

\begin{eqnarray}\label{FFGreendxyzn}
 \frac{\partial^{(n)} G({\bf r})}{\partial \tau^{(n)} } = \left(- j
k_\tau\right)^{n} \frac{e^{-j k | {\bf r} |}}{4 \pi |{{\bf r}|}},
\hspace{0.3in} \tau = x, y, z.
\end{eqnarray}

 It is not difficult to see that the far-fields and the 2D Fourier spectra
are closely related to each other.

\subsection{The 2D Fourier spectra of  the 3D spatial convolutions}\label{convolution}

It is not difficult to see that, the radiation integral in
(\ref{radiation}) can be expressed as the sum of the 3D spatial
convolutions of some source terms   with the Green's function
related expressions. For simplicity, let's consider  the 3D spatial
convolution of an arbitrary source term  ${\bf s}$ with the scalar
Green's function $G$,

\vspace{-0.1in}

\begin{eqnarray}\label{conv}
   {\bf s}({\bf r})    \stackrel{\hbox{\tiny 3D}}{\bigotimes}   G({\bf r})
    =   \oiint S  {\bf s}({\bf r}') G({\bf R})  \ dS',
 \end{eqnarray}

Now, apply the 2D Fourier transform on (\ref{conv}) and express the
scalar Green's function  $G({\bf R})$ in the spectral domain,

\vspace{-0.1in}
\begin{eqnarray}\label{convder}
\bf{\mathcal{S}}(k_x,k_y) \equiv \hbox{\large FT}_{\hbox{\tiny 2D}}
  \left[\hspace{-0.05in} \begin{array}{cccc}  \\
 \end{array}    {\bf s}({\bf r})   \stackrel{\hbox{\tiny 3D}}{\bigotimes}
  G({\bf r})  \hspace{-0.05in} \begin{array}{cccc}  \\
 \end{array}
 \right]
  \end{eqnarray}

\vspace{-0.1in}

\begin{eqnarray}
  =  \frac{1}{2 \pi }  \int_{x=-\infty}^\infty
 \int_{y=-\infty}^\infty    \left\{   \hspace{-0.05in} \begin{array}{cccc}
 \\ \\
 \end{array}    \  e^{j k_x x}  e^{j k_y
 y}    \oiint S  \left[  \hspace{-0.05in} \begin{array}{cccc}
 \\ \\
 \end{array}
     {\bf s}({\bf r}')    \times  \left( \frac{1}{2 \pi } \int_{k_x'=-\infty}^\infty
 \int_{k_y'=-\infty}^\infty  \begin{array}{cccc}
 \\ \\
 \end{array}     e^{-j k_x' (x-x')}  e^{-j
 k_y'
 (y-y')}  \right.  \right. \right. \nonumber
 \end{eqnarray}

\vspace{-0.1in}

\begin{eqnarray}
  \hspace{0.6in} \left.    \left. \left.    \times   \frac{-j e^{-j k_z' (z-z')}}{4 \pi k_z' }
    dk_x' dk_y' \hspace{-0.05in} \begin{array}{cccc}
 \\ \\
 \end{array} \right) \right]   dS'    \right\}  dx dy,        \nonumber
\end{eqnarray}

 First, do the
integral over $(x, y)$, (\ref{convder}) reduces to

\vspace{-0.1in}

\begin{eqnarray}\label{convder1}
\bf{\mathcal{S}}(k_x,k_y) =   \oiint S  \left\{ \hspace{-0.05in}
\begin{array}{cccc}
 \\ \\
 \end{array}   {\bf s}({\bf r}')    \int_{k_x'=-\infty}^\infty
 \int_{k_y'=-\infty}^\infty  \left[   e^{j k_x' x'}  e^{j
 k_y' y'}   \right.  \right.
 \end{eqnarray}

 \vspace{-0.1in}

\begin{eqnarray}
 \left.  \left.  \begin{array}{cccc}
 \\ \\
 \end{array} \times   \frac{-j e^{-j k_z' (z-z')}}{4 \pi k_z' }   \delta(k_x'-k_x)  \delta(k_y'-k_y)  dk_x' dk_y' \right]     \right\}    dS',      \nonumber
\end{eqnarray}

Next, do the integral over ($k_x'$, $k_y'$) and (\ref{convder1})
reduces to

\vspace{-0.1in}

\begin{eqnarray}\label{convder2}
\hbox{\large FT}_{\hbox{\tiny 2D}}
  \left[\hspace{-0.05in} \begin{array}{cccc}  \\
 \end{array}    {\bf s}({\bf r})   \stackrel{\hbox{\tiny 3D}}{\bigotimes}
  G({\bf r})  \hspace{-0.05in} \begin{array}{cccc}  \\
 \end{array}
 \right]   =      {\bf L}\left( \hspace{-0.05in}
\begin{array}{cccc}
 \\ \\  \end{array}  {\bf s}\left(  {\bf r} \right)\right) {\mathcal{G}^>}(k_x,k_y,0),
 \end{eqnarray}

\hspace{-0.22in}where ${\bf L}$ in (\ref{convder2}) is the radiation
vector   \cite{Balanis} for  source term ${\bf s}$, which is defined
as

\vspace{-0.1in}

\begin{eqnarray}\label{quasi-spectrum}
{\bf L}\left( \hspace{-0.05in}
\begin{array}{cccc}
 \\ \\  \end{array}  {\bf s}\left(  {\bf r} \right)\right)   =    \int \!\!\!\int_S
\hspace{-0.05in}
\begin{array}{cccc}
 \\ \\  \end{array}   {\bf s}({\bf r}')
  e^{j {\bf k} \cdot {\bf r}'} dS',
\end{eqnarray}

It is not difficult to see that  the radiation vector ${\bf L}$ in
(\ref{quasi-spectrum}) reduces to the regular 2D Fourier spectrum
when   surface $S$ is a plane   located at $z'=0$.

\vspace{-0.1in}

\begin{eqnarray}\label{quasi-spectrum1}
\left.{\bf L}\left( \hspace{-0.05in}
\begin{array}{cccc}
 \\ \\  \end{array}  {\bf s}\left(  {\bf r} \right)\right)\right|_{z'=0}   = 2 \pi   \hbox{\large FT}_{\hbox{\tiny 2D}}   \left[\hspace{-0.05in} \begin{array}{cccc}  \\
 \end{array}   {\bf s}(x, y)  \hspace{-0.05in} \begin{array}{cccc}
 \\ \\
 \end{array}
 \right],
\end{eqnarray}

\hspace{-0.22in}where the dummy primed ($x',y'$) have been replaced
with ($x,y$). Substitute (\ref{quasi-spectrum}) into
(\ref{convder2}),

\vspace{-0.1in}

\begin{eqnarray}\label{convder3}
 \bf{\mathcal{S}}(k_x,k_y)  =     \frac{-j e^{-j k_z z}}{4 \pi k_z }
 {\bf L}\left( \hspace{-0.05in}
\begin{array}{cccc}
 \\ \\  \end{array}  {\bf s}\left(  {\bf r} \right)\right) =   \mathcal{G}(k_x,k_y,0)    {\bf L}\left( \hspace{-0.05in}
\begin{array}{cccc}
 \\ \\  \end{array}  {\bf s}\left(  {\bf r} \right)\right).
 \end{eqnarray}

It is not difficult to see that the radiation vector ${\bf L}$ in
(\ref{quasi-spectrum}) is closely related to the far-field   by
letting ${\bf r} \rightarrow \infty$,

\vspace{-0.1in}

\begin{eqnarray}\label{convFF}
   \left.{\bf s}({\bf r})   \stackrel{\hbox{\tiny 3D}}{\bigotimes}    G({\bf
   r})\right|_{{\bf r} \rightarrow \infty}
    =  \frac{e^{-j k |{\bf r}|}}{ 4 \pi |{\bf r}|}  {\bf L}\left( \hspace{-0.05in}
\begin{array}{cccc}
 \\ \\  \end{array}  {\bf s}\left(  {\bf r} \right)\right).
 \end{eqnarray}

From (\ref{convFF}), if   $\frac{e^{-j k |{\bf r}|}}{ 4 \pi |{\bf
r}|}$ can be  ignored, the radiation vector ${\bf L}$ can be
considered as  the far-field pattern, which means that when the
far-field is obtained, the radiation vector and the 2D Fourier
spectrum of the 3D convolution are also obtained, from
(\ref{convFF}) and (\ref{convder3}) respectively.

\subsection{2D Fourier spectrum of the radiation integral}\label{2DFTradiation}

From (\ref{convder3}) and (\ref{radiation}),  the 2D Fourier
spectrum of  the radiation integral (denoted as $ \bf{\mathcal{F}}$)
is obtained, which is

\begin{eqnarray}\label{Kottler2DFT}
 \hspace{-0.2in} \bf{\mathcal{F}}   =
  \frac{-j}{ \omega \epsilon}  {\mathcal{G}^>}(k_x,k_y,0) \left\{  \hspace{-0.05in} \begin{array}{cccc} \\
  \\  \\
 \end{array}  k^2
  {\bf L} \left(\hspace{-0.05in} \begin{array}{cccc}  \\ \\
 \end{array}  {\bf J}({\bf r})  \hspace{-0.05in} \begin{array}{cccc}  \\
 \end{array}
 \right)
- {\bf k} \sum_{\tau=x,y,z} \left[  \hspace{-0.05in}
\begin{array}{cccc}  \\  \\
 \end{array} k_{\tau} {\bf L} \left(\hspace{-0.05in} \begin{array}{cccc}
 \\  \\
 \end{array}  {\bf J}_{\tau}({\bf r})   \hspace{-0.05in} \begin{array}{cccc}  \\
 \end{array}
 \right)  \right]   +  \omega \epsilon  {\bf L} \left( \hspace{-0.05in} \begin{array}{cccc}
 \\  \\
 \end{array}  {\bf J}_m({\bf r})   \hspace{-0.05in} \begin{array}{cccc}  \\
 \end{array}
 \right)    \times {\bf k} \hspace{-0.05in} \begin{array}{cccc}  \\
 \end{array} \right\}.
\end{eqnarray}

\subsection{Electromagnetic field on a plane}\label{subsec:pws}

After the 2D Fourier spectrum $ \bf{\mathcal{F}}$ has been obtained,
the electric field $\hbox{\bf E}$ can be expressed in the PWS form
\cite{Whittaker,Booker}, which is given as,

  \vspace{-0.1in}

\begin{eqnarray}\label{PWS}
 \hbox{\bf E}({\bf r})  =   \hbox{\large IFT}_{\hbox{\tiny 2D}}  \left[  \hspace{-0.05in} \begin{array}{cccc}  \\  \\  \end{array}
 \bf{\mathcal{F}}_0
(k_x,k_y) e^{-j k_z z}  \hspace{-0.05in} \begin{array}{cccc}  \\  \\
\end{array} \right]
\end{eqnarray}

 \hspace{-0.22in}where $ \bf{\mathcal{F}}_0$ and the 2D Inverse Fourier
Transform have been defined as

\vspace*{-0.1in}

\begin{eqnarray}
 \bf{\mathcal{F}}_0
(k_x,k_y)    =  {\mathcal{F}}_{0x} \hat{ \bf x} + {\mathcal{F}}_{0y}
\hat{ \bf y}  +  {\mathcal{F}}_{0z} \nonumber
   \hat{ \bf z}  =   \bf{\mathcal{F}} (k_x,k_y)  e^{j k_z z}, \nonumber
   \end{eqnarray}

\vspace*{-0.1in}

\begin{eqnarray}
 {\mathcal{F}}_{0z} = - \frac{k_x
  {\mathcal{F}}_{0x} + k_y  {\mathcal{F}}_{0y}}{k_z},  \nonumber
\end{eqnarray}

   \vspace*{-0.1in}

 \begin{eqnarray}\label{2DIFTdef}
 \hspace*{-0.2in} \hbox{\large IFT}_{\hbox{\tiny 2D}}  \left[\hspace{-0.05in} \begin{array}{cccc}  \\
 \end{array} \cdot  \hspace{-0.05in} \begin{array}{cccc}  \\
 \end{array}
 \right]  =  \frac{1}{2 \pi} \int_{k_x=-\infty}^\infty  \left\{    e^{-j k_x
 x} \int_{k_y=-\infty}^\infty
\left[\hspace{-0.05in} \begin{array}{cccc}  \\
 \end{array}  \cdot    \hspace{-0.05in} \begin{array}{cccc}  \\
 \end{array} \right]    e^{-j k_y y} dk_y   \right\} dk_x. \nonumber
\end{eqnarray}

\section{The Planar TI-FFT Algorithm}\label{sec:TIFFT}

In this section, the optimized spatial and spectral TI-FFTs are
presented. It will be shown that both of them have the same
computational complexity for the same quasi-planar surface.

\subsection{The {spatial} TI-FFT algorithm}\label{spatialTIFFT}

It has been shown in (\ref{PWS}) that the electric field ${\bf E}$
on a plane can be evaluated through the 2D inverse Fourier
transform. For a quasi-planar surface, the TI technique can  be
used, which leads to the {spatial} TI-FFT  algorithm (where the
quasi-planar surface is sliced into many small spatial subdomains,
as shown Fig. \ref{fig:scheme}).

Rewrite the electric field $  {\bf E}$ in (\ref{PWS}) as follows,

\vspace*{-0.1in}

\begin{eqnarray}\label{spacedivision_spa}
 {\bf E}({\bf r})  =  e^{-j k z}  {\hbox{\large IFT}}_{\hbox{\tiny 2D}}
   \left[\hspace{-0.05in} \begin{array}{cccc}  \\  \\
 \end{array}   \left[ \hspace{-0.05in} \begin{array}{cccc}  \\
 \end{array}  \bf{  \widetilde{\mathcal{F}}}_0(k_x, k_y)    \hspace{-0.05in} \begin{array}{cccc}
 \\
 \end{array}  \right] e^{j \triangle k_z  \triangle z}  \hspace{-0.05in} \begin{array}{cccc}  \\ \\
 \end{array}
 \right],  \ \ \ \
\bf{  \widetilde{\mathcal{F}}}_0(k_x, k_y) =   \bf{
\mathcal{F}}_0(k_x, k_y) e^{j  \triangle k_z   z_{\hbox{\tiny
  min}}}
   \end{eqnarray}

\hspace{-0.2in}where  $z_{\hbox{\tiny min}}$ denotes the minimum
value of $z$. Now, express $e^{-j \triangle k_z \triangle z}$ into a
Taylor series on the spatial reference plane located at $z = z_r$,

\vspace*{-0.1in}

\begin{eqnarray}\label{space_sliceTI}
e^{j \triangle k_z \triangle z} =  e^{j \triangle k_z \triangle z_r}
\sum_{n=0}^{\mathcal{N}_o}\left[ \frac{1}{n!} \left( j \triangle k_z
\right)^{n} \left( z  - z_r \hspace{-0.05in}
\begin{array}{cccc}
 \\   \end{array}\right)^{n}
\right],
\end{eqnarray}

\hspace{-0.22in}where $ \triangle z_r = z_r - z_{\hbox{\tiny min}}$
and $\mathcal{N}_o$ is the order of Taylor series.

Substitute (\ref{space_sliceTI}) into (\ref{spacedivision_spa}), the
spatial TI-FFT  algorithm  for the electric field ${\bf
 E}$ is obtained,

\vspace{-0.1in}

\begin{eqnarray}\label{spatialdivisionTIFFT}
 {\bf E}({\bf r})       = e^{-j k z}
 \sum_{n=0}^{\mathcal{N}_o} \left\{  \hspace{-0.05in} \begin{array}{cccc}
 \\  \\
 \end{array}  \frac{1}{n!}  \left[ j \left(   z  - z_r
 \right)   \hspace{-0.05in}
\begin{array}{cccc}
 \\  \\  \end{array}  \right]^n    \hbox{\large IFT}_{\hbox{\tiny 2D}}   \left[\hspace{-0.05in} \begin{array}{cccc}  \\
 \end{array}     \bf{  \widetilde{\mathcal{F}}}_0(k_x, k_y)   e^{j \triangle k_z \triangle z_r} \left(    \hspace{-0.05in}
\begin{array}{cccc}
 \\   \end{array}  \triangle k_z   \right)^{n} \hspace{-0.05in} \begin{array}{cccc}
 \\
 \end{array}
 \right]   \hspace{-0.05in} \begin{array}{cccc}
 \\  \\
 \end{array}\right\},
\end{eqnarray}

The number of spatial reference planes $\mathcal{N}_r$ required in
the computation  depends on the spatial slicing spacing
$\left(\delta_z \equiv \hbox{max}
\left[\hspace{-0.05in} \begin{array}{cccc}  \\
 \end{array}  z  -
z_r \hspace{-0.05in} \begin{array}{cccc}  \\
 \end{array} \right] = z_{r+1} - z_r\right)$  and the characteristic surface variation $\triangle z_c$ (within
which the electromagnetic field is of interest): $\mathcal{N}_r
\propto \triangle z_c / \delta_z $. The readers should note that the
actual  maximum interpolation distance
 is $\frac{\delta_z}{2}$, which is located at the middle of two adjacent spatial
 reference planes, but $\delta_z$ is used in this article to simplify the notation.  Apparently, to achieve the
desired computational accuracy (denoted as $\gamma_{\hbox{\tiny
TI}}$), the choice of the spatial slicing spacing $\delta_z$ between
two adjacent spatial reference planes depends on $\triangle
k_{z,c}$, which is  defined as

\vspace{-0.1in}

\begin{eqnarray}\label{kzc}
 \triangle k_{z,c} \equiv k - k_{z,c} = k - \sqrt{k^2 -
 k_{\perp,c}^2} = k \alpha,  \ \ \ \ \alpha =  1 - \sqrt{1 -
 \left(\frac{k_{\perp,c}}{k}\right)^2},
\end{eqnarray}

\hspace{-0.22in}where, $k_{\perp,c}$ is the characteristic bandwidth
(beyond which the 2D Fourier spectrum $\bf{\mathcal{F}}$ is
negligible) and is defined on x-y plane.  It is clear that the
smaller the bandwidth $k_{\perp,c}$, the larger the $\delta_z$ could
be, which also means a smaller $\mathcal{N}_r$. So, a narrow-band
beam and a small surface variation $\triangle z_c$ (quasi-planar
geometry) are in favor of the planar TI-FFT algorithm.

\begin{figure}\centering
\includegraphics[width= 5.5 in, height= 4.0 in]{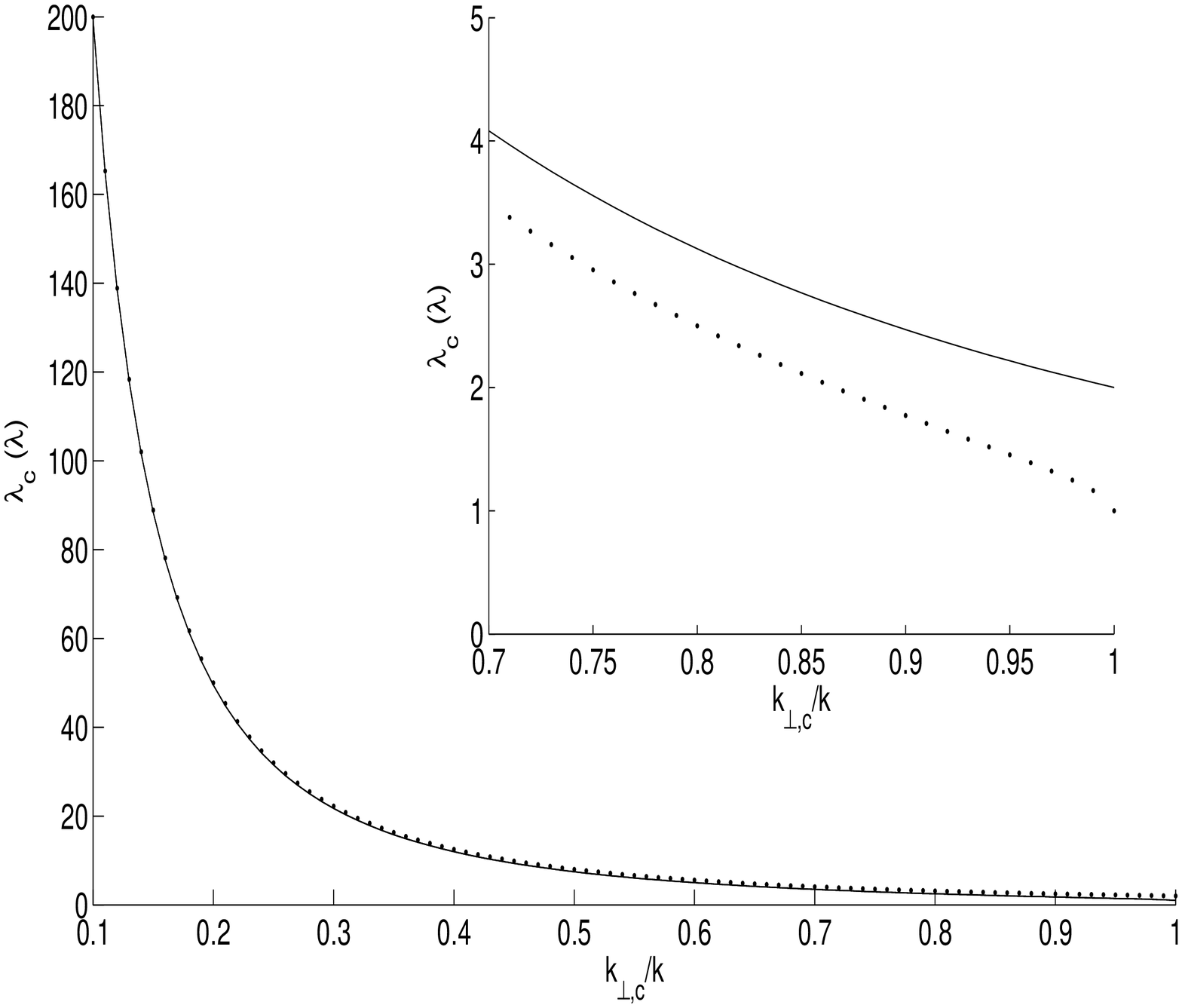}
\caption{ The plots of the characteristic wavelength $\lambda_{c}$
for different $k_{z,c}$. The exact value (line) is given in
(\ref{cwavelength}) and the approximation (dots) is given in
(\ref{cwavelength1}). The plots show  that $\lambda_c \gg \lambda$
for a narrow-band beam. The maximum deviation of  the approximation
from the exact value is $1 \lambda$, which occurs at $k_{z,c}=0$
($k_{\perp,c}=k$).}
 \label{fig:wavelength}
\end{figure}

In view of the importance of the spatial slicing spacing $\delta_z$,
it is helpful to define the characteristic wave length $\lambda_{c}
$ for a narrow-band beam. From (\ref{kzc}),

\vspace{-0.1in}

\begin{eqnarray}\label{cwavelength}
  \lambda_{c} \equiv \frac{2\pi}{ \triangle k_{z,c}} =  \frac{2\pi}{k - \sqrt{k^2 - k_{\perp,c}^2}}
  =  \frac{\lambda}{\alpha},
\end{eqnarray}

For a narrow-band beam ($k_{\perp,c} \ll k$),

\vspace{-0.1in}

\begin{eqnarray}\label{cwavelength1}
  \lambda_{c} \sim 2 \left(\frac{k}{k_{\perp,c}}\right)^2 \lambda.
\end{eqnarray}

  Fig. \ref{fig:wavelength} plots the exact value in (\ref{cwavelength})  and
approximation  in (\ref{cwavelength1}) of the characteristic
wavelength $\lambda_{c}$ for different characteristic bandwidth
$k_{\perp,c}$, from which it can be seen that the maximum deviation
of the approximation from the exact value is
  $1 \lambda$, which occurs at $k_{\perp,c} = k$.

It can be seen from (\ref{space_sliceTI}) and
(\ref{spatialdivisionTIFFT}) that,  for the given computational
accuracy $\gamma_{\hbox{\tiny TI}}$, the spatial slicing spacing
$\delta_z$ should satisfy the following relation,

\vspace{-0.1in}

\begin{eqnarray}\label{accuracy}
   \gamma_{\hbox{\tiny TI}}   \sim \mathcal{O} \left[ \hspace{-0.05in} \begin{array}{cccc}  \\
 \end{array} \left(\triangle k_{z,c} \ \delta_z\right)^{\mathcal{N}_o}  \hspace{-0.05in} \begin{array}{cccc}  \\
 \end{array}\right]
\end{eqnarray}

  \vspace{-0.11in}

\begin{eqnarray}\label{deltazexpr}
\rightarrow  \delta_z \sim  \frac{1}{k \alpha} \left( \hspace{-0.05in} \begin{array}{cccc}  \\
 \end{array} \frac{1}{\gamma_{\hbox{\tiny TI}}} \hspace{-0.05in} \begin{array}{cccc}  \\
 \end{array} \right)^{-\frac{1}{\mathcal{N}_o}}   =  \frac{\lambda}{2
 \pi \alpha}   \left( \hspace{-0.05in} \begin{array}{cccc}  \\
 \end{array} \frac{1}{\gamma_{\hbox{\tiny TI}}} \hspace{-0.05in} \begin{array}{cccc}  \\
 \end{array} \right)^{-\frac{1}{\mathcal{N}_o}},
\end{eqnarray}

 For a narrow-band beam ($k_{\perp,c} \ll k$),

\vspace{-0.1in}

\begin{eqnarray}\label{deltazexpr1}
 \delta_z    \sim \frac{1}{\pi} \left(\frac{k}{k_{\perp,c}}\right)^2
 \left( \hspace{-0.05in} \begin{array}{cccc}  \\
 \end{array} \frac{1}{\gamma_{\hbox{\tiny TI}}} \hspace{-0.05in} \begin{array}{cccc}  \\
 \end{array} \right)^{-\frac{1}{\mathcal{N}_o}}  \lambda.
\end{eqnarray}

Now consider a quasi-planar surface with a characteristic surface
variation of $\triangle z_c = \mathcal{N}_z \lambda$, from
(\ref{deltazexpr}) the number of spatial reference planes
$\mathcal{N}_r$ is given as

\vspace{-0.1in}

\begin{eqnarray}\label{Nr}
 \mathcal{N}_r = \frac{\triangle z_c}{\delta_z}  \sim 2 \pi \alpha \left( \hspace{-0.05in} \begin{array}{cccc}  \\
 \end{array} \frac{1}{\gamma_{\hbox{\tiny TI}}} \hspace{-0.05in} \begin{array}{cccc}  \\
 \end{array} \right)^{\frac{1}{ \mathcal{N}_o}}  \mathcal{N}_z,
\end{eqnarray}

For a narrow-band beam ($k_{\perp,c} \ll k$),

\vspace{-0.1in}

\begin{eqnarray}\label{Nr1}
 \mathcal{N}_r  \sim \pi
   \left(\frac{k_{\perp,c} }{k }\right)^2
 \left( \hspace{-0.05in} \begin{array}{cccc}  \\
 \end{array} \frac{1}{\gamma_{\hbox{\tiny TI}}} \hspace{-0.05in} \begin{array}{cccc}  \\
 \end{array} \right)^{\frac{1}{ \mathcal{N}_o}}  \mathcal{N}_z.
\end{eqnarray}

The number of FFT operations $N_{\hbox{\tiny FFT}}$ and the
computational complexity $\hbox{CPU}$ are obtained as

\vspace{-0.1in}

\begin{eqnarray}\label{NFFT}
N_{\hbox{\tiny FFT}} = \mathcal{N}_o \times \mathcal{N}_r =  2\pi \alpha \left( \hspace{-0.05in} \begin{array}{cccc}  \\
 \end{array} \frac{1}{\gamma_{\hbox{\tiny TI}}} \hspace{-0.05in} \begin{array}{cccc}  \\
 \end{array} \right)^{\frac{1}{ \mathcal{N}_o}}   \mathcal{N}_o
 \mathcal{N}_z,
\end{eqnarray}

\vspace{-0.1in}

\begin{eqnarray}\label{complexity}
\hbox{CPU}  =  N_{\hbox{\tiny FFT}}  \ \mathcal{O} \left[\hspace{-0.05in} \begin{array}{cccc}  \\
 \end{array} N \log_2
N \hspace{-0.05in} \begin{array}{cccc}  \\
 \end{array} \right]   = 2 \pi \alpha \left( \hspace{-0.05in} \begin{array}{cccc}  \\
 \end{array} \frac{1}{\gamma_{\hbox{\tiny TI}}} \hspace{-0.05in} \begin{array}{cccc}  \\
 \end{array} \right)^{\frac{1}{ \mathcal{N}_o}}   \mathcal{N}_o  \mathcal{N}_z \ \mathcal{O}  \left[\hspace{-0.05in} \begin{array}{cccc}  \\
 \end{array} N \log_2 N  \hspace{-0.05in} \begin{array}{cccc}  \\
 \end{array}\right],
 \end{eqnarray}

For a narrow-band beam ($k_{\perp,c} \ll k$),

\vspace{-0.1in}

\begin{eqnarray}\label{NFFT1}
N_{\hbox{\tiny FFT}} \sim \pi
   \left(\frac{k_{\perp,c} }{k }\right)^2
 \left( \hspace{-0.05in} \begin{array}{cccc}  \\
 \end{array} \frac{1}{\gamma_{\hbox{\tiny TI}}} \hspace{-0.05in} \begin{array}{cccc}  \\
 \end{array} \right)^{\frac{1}{ \mathcal{N}_o}}   \mathcal{N}_o
 \mathcal{N}_z.
\end{eqnarray}

\vspace{-0.1in}

\begin{eqnarray}\label{complexity1}
 \hbox{CPU}   \sim  \pi
   \left(\frac{k_{\perp,c} }{k }\right)^2
 \left( \hspace{-0.05in} \begin{array}{cccc}  \\
 \end{array} \frac{1}{\gamma_{\hbox{\tiny TI}}} \hspace{-0.05in} \begin{array}{cccc}  \\
 \end{array} \right)^{\frac{1}{ \mathcal{N}_o}}   \mathcal{N}_o  \mathcal{N}_z \ \mathcal{O}  \left[\hspace{-0.05in} \begin{array}{cccc}  \\
 \end{array} N \log_2 N  \hspace{-0.05in} \begin{array}{cccc}  \\
 \end{array}\right].
\end{eqnarray}

 For a narrow-band beam, the
computational complexity $\hbox{CPU}$ has a square law dependence on
the characteristic bandwidth $k_{\perp,c}$ of the electromagnetic
wave and have a linear dependence on the surface variation
($\triangle z_c = \mathcal{N}_z\lambda$). The computational
complexity $\hbox{CPU}$ also has an inverse ${
\mathcal{N}_o}^{\hbox{\tiny th}}$-root dependence on the
computational accuracy $\gamma_{\hbox{\tiny TI}}$. So   the
characteristic bandwidth $k_{\perp,c}$  has the most significant
effect on the computational complexity of the planar TI-FFT
algorithm.

\begin{figure}\centering
\includegraphics[width= 5.5 in, height= 4.0 in]{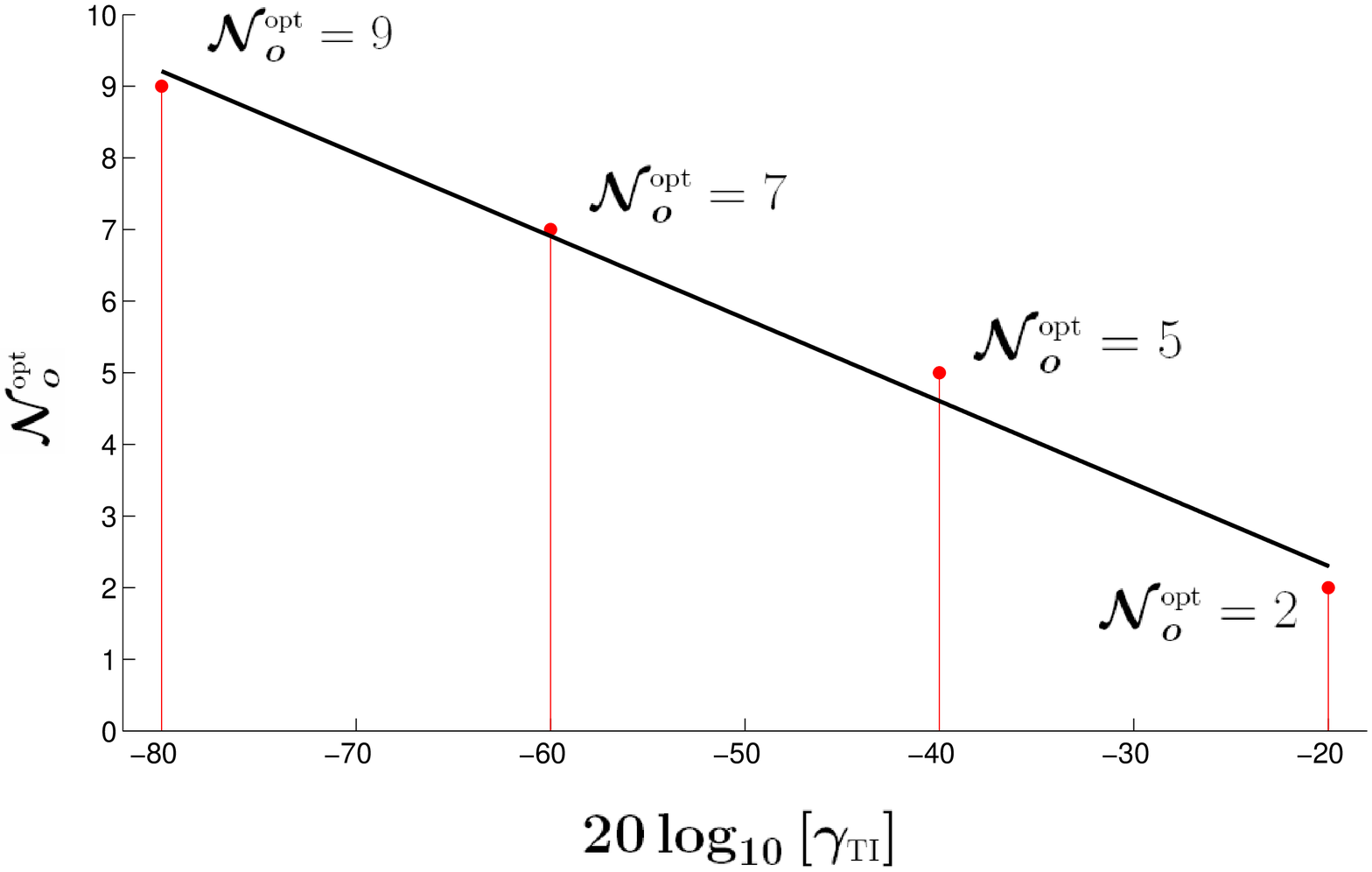}
\caption{The linear dependence of the optimized order of Taylor
series $\mathcal{N}_o^{\hbox{\tiny opt}}$ on the computational
accuracy $\gamma_{\hbox{\tiny
    TI}}$ (dB). It can be seen that $\mathcal{N}_o^{\hbox{\tiny opt}}
    = (2, 5, 7, 9)$ for $\gamma_{\hbox{\tiny
    TI}} = (-20, -40, -60, -80)$ dB respectively. }
 \label{fig:Nt}
\end{figure}

It can be seen from (\ref{NFFT}) or (\ref{complexity}) that  the
optimized number of Taylor series $\mathcal{N}_o^{\hbox{\tiny opt}}$
can  be obtained  through finding the minimum value of
$N_{\hbox{\tiny FFT}}$ in (\ref{NFFT}) or CPU in (\ref{complexity})
by assuming that $\mathcal{N}_o$ is a continuous variable,

\vspace{-0.1in}

\begin{eqnarray}\label{Optimization}
 \left.\frac{\partial N_{\hbox{\tiny FFT}}}{\partial \mathcal{N}_o}\right|_{\mathcal{N}_o^{\hbox{\tiny opt}}}   = 0
\rightarrow
   \left.\frac{\partial  \left[  \ln\left[\mathcal{N}_o\right] - \ln \left[\gamma_{\hbox{\tiny TI}}\right]  \frac{1}{\mathcal{N}_o}
 \right]
 } {\partial \mathcal{N}_o}\right|_{\mathcal{N}_o^{\hbox{\tiny opt}}} =
 0,
  \end{eqnarray}

\vspace{-0.1in}

\begin{eqnarray}\label{No1}
  \mathcal{N}_o^{\hbox{\tiny opt}}   \sim  \hbox{round} \left[ \hspace{-0.09in} \begin{array}{cccc}
  \\ \\
 \end{array}   \ln \left[ \frac{1}{\gamma_{\hbox{\tiny
    TI}}} \hspace{-0.05in} \begin{array}{cccc}
  \\
 \end{array} \right] \right]  =   \hbox{round} \left[\hspace{-0.05in} \begin{array}{cccc}  \\
 \end{array} - 0.1151 \gamma_{\hbox{\tiny TI}}
    (\hbox{dB})  \hspace{-0.05in} \begin{array}{cccc}  \\  \\
 \end{array} \right].
\end{eqnarray}

\hspace{-0.22in}where ``round" means to round the value to its
nearest integer (actually, to achieve a higher computational
accuracy $\gamma_{\hbox{\tiny TI}}$, the upper-bound could be used
but the computational complexity CPU  is a little higher). Fig.
\ref{fig:Nt} also shows the linear dependence of the optimized order
of Taylor series $\mathcal{N}_o^{\hbox{\tiny opt}}$ on the
computational accuracy $\gamma_{\hbox{\tiny TI}}$ (dB),  which has
been shown in (\ref{No1}).

\begin{figure}[t]\centering
 \includegraphics[width= 5.5 in, height= 4.0 in]{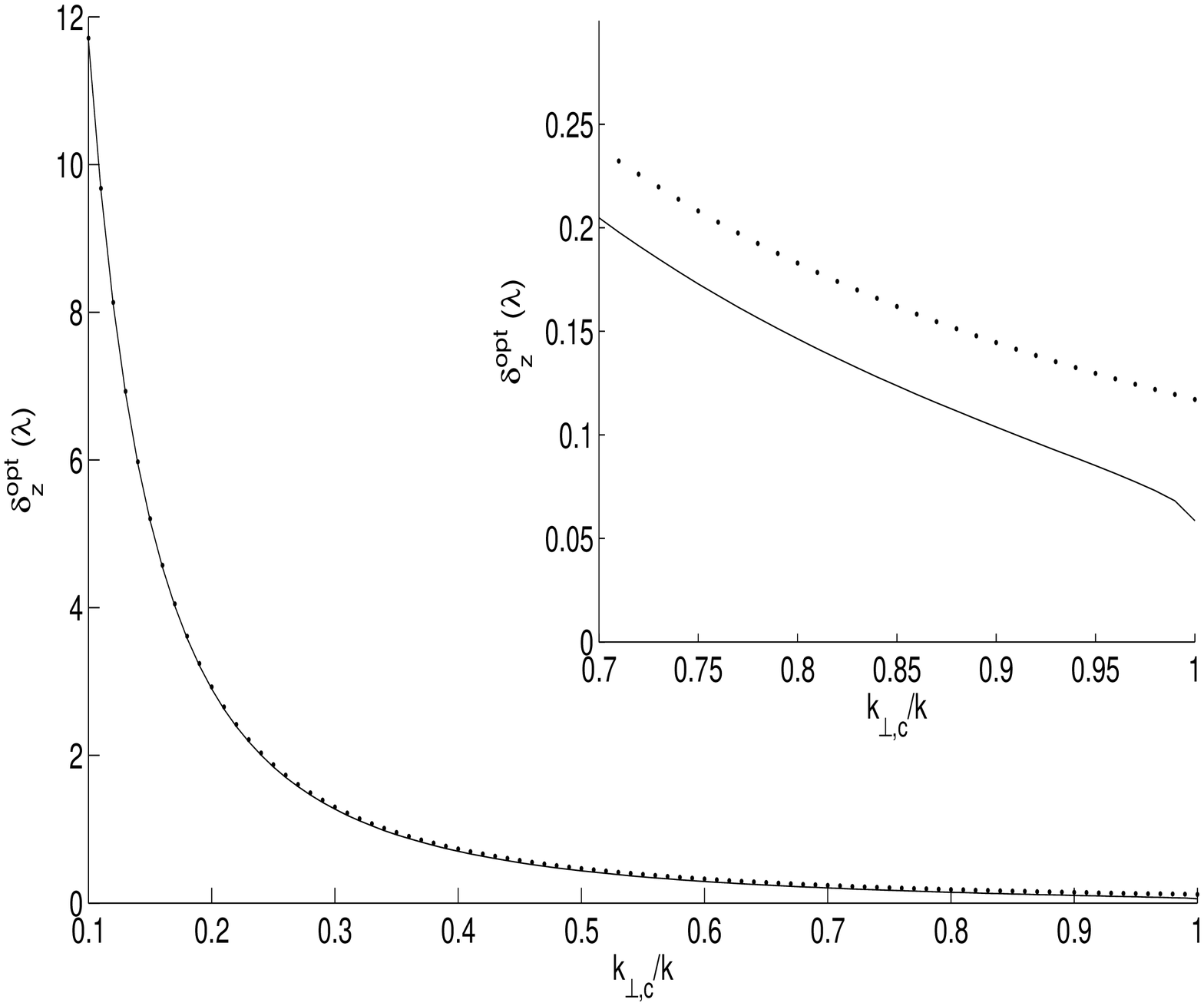}
\caption{ Plots of  the exact value (line)  of
$\delta_z^{\hbox{\tiny opt}}$ given in (\ref{deltazexpropt})   and
approximation (dots) given in (\ref{deltazexpropt1})
  for different  characteristic bandwidth $k_{\perp,c}$.}
 \label{fig:deltazopt}
\end{figure}

The optimized spatial slicing spacing $\delta_z^{\hbox{\tiny opt}} $
can be obtained from (\ref{deltazexpr}) and (\ref{No1}), which is

\vspace{-0.1in}

\begin{eqnarray}\label{deltazexpropt}
  \delta_z^{\hbox{\tiny opt}} \sim \frac{\lambda}{2
 \pi \alpha}   \left( \hspace{-0.05in} \begin{array}{cccc}  \\
 \end{array} \frac{1}{\gamma_{\hbox{\tiny TI}}} \hspace{-0.05in} \begin{array}{cccc}  \\
 \end{array} \right)^{-\frac{1}{\ln \left[ \frac{1}{\gamma_{\hbox{\tiny
    TI}}}\right]}}  = \frac{\lambda}{2
 \pi e \alpha}   \sim \frac{1}{17} \lambda_c,
\end{eqnarray}

For a narrow-band beam ($k_{\perp,c} \ll k$),

\vspace{-0.1in}

\begin{eqnarray}\label{deltazexpropt1}
  \delta_z^{\hbox{\tiny opt}}   \sim \frac{1}{e\pi} \left(\frac{k}{k_{\perp,c}}\right)^2
  \lambda,
\end{eqnarray}

\hspace{-0.22in}where $e \sim 2.718$ is the natural logarithmic
base. It is interesting to note that the optimized spatial slicing
spacing $\delta_z$ doesn't depend on the computational accuracy
$\gamma_{\hbox{\tiny TI}}$ and strongly depends on the
characteristic bandwidth $k_{\perp,c}$ (inverse square law). Fig.
\ref{fig:deltazopt} shows $\delta_z^{\hbox{\tiny opt}}$ for
different characteristic bandwidth $k_{\perp,c}$, from which it can
be seen that $\delta_z^{\hbox{\tiny opt}} > 0.5 \lambda$ for
$k_{z,c}
> 0.9 k$ ($k_{\perp,c} < 0.436 k$).

The optimized number of spatial reference planes $
\mathcal{N}_r^{\hbox{\tiny opt}} $  is given as

\vspace{-0.1in}

\begin{eqnarray}\label{Nropt}
 \mathcal{N}_r^{\hbox{\tiny opt}}  = \frac{\triangle z_c}{\delta_z}
 = 2 \pi e \alpha \mathcal{N}_z \sim 17 \alpha  \mathcal{N}_z,
\end{eqnarray}

For a narrow-band beam ($k_{\perp,c} \ll k$),

\vspace{-0.1in}

\begin{eqnarray}\label{Nropt1}
 \mathcal{N}_r^{\hbox{\tiny opt}}    \sim
 \pi e
   \left(\frac{k_{\perp,c} }{k }\right)^2
 \mathcal{N}_z.
\end{eqnarray}

 Substitute (\ref{No1}) into (\ref{NFFT}) and
(\ref{complexity}), the optimized number of  FFT operations
$N_{\hbox{\tiny FFT}}^{\hbox{\tiny opt}}$ and the optimized
computational complexity $ \hbox{CPU}^{\hbox{\tiny opt}}$ can also
be obtained,

\vspace{-0.1in}

\begin{eqnarray}\label{NFFTopt}
N_{\hbox{\tiny FFT}}^{\hbox{\tiny opt}}  = 2 \pi e \alpha  \ln
\left[ \frac{1}{\gamma_{\hbox{\tiny TI}}}\right] \mathcal{N}_z,
\end{eqnarray}

For a narrow-band beam ($k_{\perp,c} \ll k$),

\vspace{-0.1in}

\begin{eqnarray}\label{NFFToptapp}
N_{\hbox{\tiny FFT}}^{\hbox{\tiny opt}}  \sim \pi e
   \left(\frac{k_{\perp,c} }{k }\right)^2
   \ln \left[ \frac{1}{\gamma_{\hbox{\tiny TI}}}\right]
   \mathcal{N}_z.
\end{eqnarray}

The optimized computational complexity $ \hbox{CPU}^{\hbox{\tiny
opt}}$ is given as

\begin{eqnarray}\label{complexityopt}
\hbox{CPU}^{\hbox{\tiny opt}}  &\sim& N_{\hbox{\tiny FFT}}^{\hbox{\tiny opt}}    \mathcal{O}  \left[\hspace{-0.05in} \begin{array}{cccc}  \\
 \end{array} N \log_2 N  \hspace{-0.05in} \begin{array}{cccc}  \\
 \end{array}\right].
\end{eqnarray}

\subsection{The {spectral} TI-FFT algorithm}\label{spectralTIFFT}

\begin{figure}\centering
\includegraphics[width= 5 in, height= 5  in]{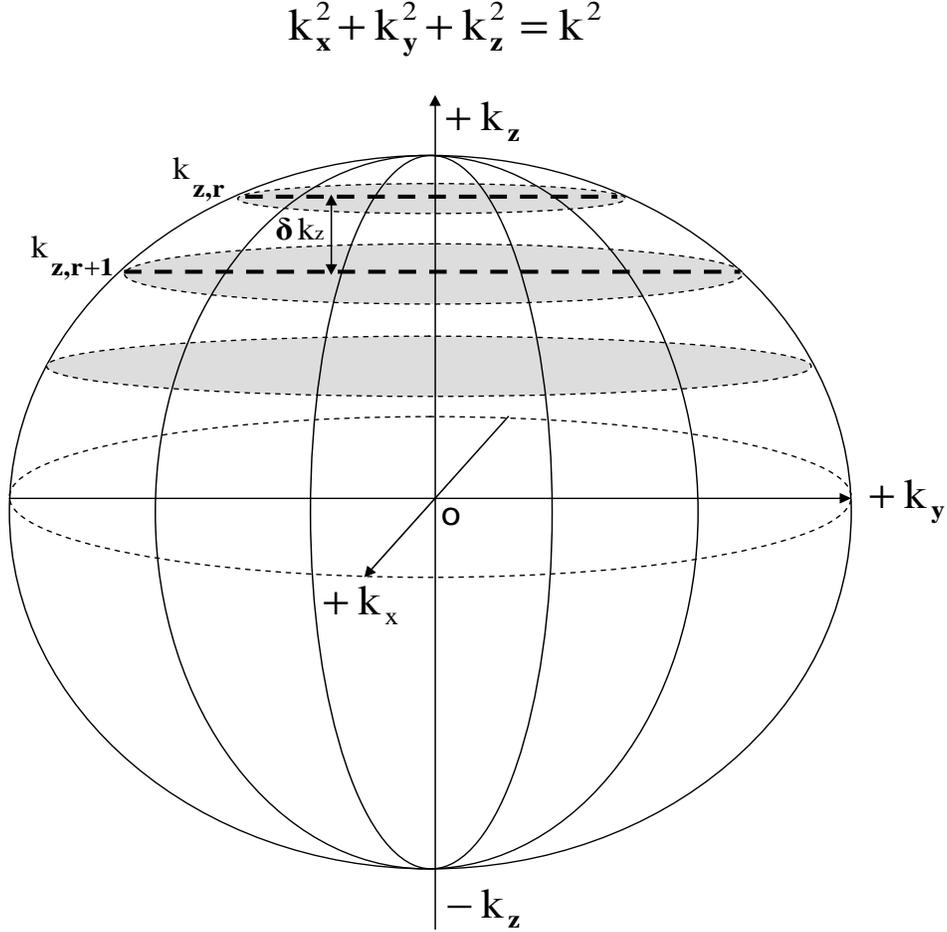}
\caption{The spectral domain division   for  the {\sl spectral}  and
TI-FFT: Only $k_z
>0$ half sphere surface is used for   half-space $z > z'$ computation in
this article.  $k_{z,r}$ and $k_{z,r+1}$ denote the $r^{\hbox{\tiny
th}}$ and $(r+1)^{\hbox{\tiny th}}$ spectral reference planes
respectively. $\delta_{k_z}$ is the spectral slicing spacing.}
 \label{fig:spectral}
\end{figure}

It has been shown in (\ref{Kottler2DFT}) that the computation of the
2D Fourier spectrum $\bf{\mathcal{F}}$ is equivalent to evaluate the
radiation vector  ${\bf L}$. For the quasi-planar geometry, the FFT
can still be used with the help of the TI technique, which leads to
the spectral TI-FFT algorithm (where the spherical spectral surface
is sliced into many small spectral subdomains, as shown Fig.
\ref{fig:spectral}).

From (\ref{quasi-spectrum}), the radiation vector ${\bf L}$ can be
rewritten as

\vspace{-0.1in}

\begin{eqnarray}\label{spectral_slice}
 {\bf L}\left({\bf f} ({\bf r}) \hspace{-0.05in} \begin{array}{cccc}  \\ \\
 \end{array} \right)  =
 2 \pi e^{j  k_z  z_{\hbox{\tiny min}}} {\hbox{\large FT}}_{\hbox{\tiny
2D}}  \left[ \hspace{-0.05in} \begin{array}{cccc}  \\
 \end{array}    \widetilde{\bf f}({\bf r})    e^{ j  k_z \triangle z}   \hspace{-0.05in} \begin{array}{cccc}  \\ \\
 \end{array}
 \right],
  \end{eqnarray}

\hspace{-0.22in}where $\widetilde{\bf f}({\bf r}) = \frac{ {{\bf
s}}({\bf r})}{{\hat{\bf n} \cdot \hat{\bf z}}}$ and $\hat{\bf n}$ is
the normal to surface $S$. Now the Taylor expansion of $ {\bf L}$ in
(\ref{spectral_slice}) over $k_z$ is given as

\vspace{-0.1in}

\begin{eqnarray}\label{spectraldivisionTIFFT}
 {\bf L}\left({\bf f} ({\bf r}) \hspace{-0.05in} \begin{array}{cccc}  \\ \\
 \end{array} \right)     =    2 \pi e^{j  k_z  z_{\hbox{\tiny min}}} \sum_{n=0}^{\mathcal{N}_o}
  \left\{ \hspace{-0.05in} \begin{array}{cccc}
 \\ \\
 \end{array}  \frac{1}{n!}  \left(\hspace{-0.05in} \begin{array}{cccc}
 \\
 \end{array} j
\left[k_z - k_{z,r}\right]  \hspace{-0.05in} \begin{array}{cccc}  \\
 \end{array} \right)^{n}       \hbox{\large FT}_{\hbox{\tiny 2D}}   \left[\hspace{-0.05in} \begin{array}{cccc}  \\
 \end{array}     \widetilde{\bf f}({\bf r})   \left( \hspace{-0.05in} \begin{array}{cccc}  \\
 \end{array}  \triangle z \hspace{-0.05in} \begin{array}{cccc}  \\
 \end{array}  \right)^n \hspace{-0.05in} \begin{array}{cccc}  \\
 \end{array}
 \right]    \hspace{-0.14in} \begin{array}{cccc}  \\  \\
 \end{array}\right\},
\end{eqnarray}

\hspace{-0.22in}where $k_{z,r}$ denotes the spectral reference
plane.  For the given computational accuracy $\gamma_{\hbox{\tiny
TI}}$, the spectral slicing spacing $\left(\delta k_z \equiv \hbox{max} \left[ \hspace{-0.05in} \begin{array}{cccc}  \\
 \end{array} k_z - k_{z,r} \hspace{-0.05in} \begin{array}{cccc}  \\
 \end{array}    \hspace{-0.05in} \begin{array}{cccc}  \\
 \end{array} \right] = k_{z,r+1} - k_{z,r}\right)$
should satisfy the following relation,

\vspace{-0.1in}

\begin{eqnarray}\label{accuracyspec}
   \gamma_{\hbox{\tiny TI}}   \sim \mathcal{O} \left[ \hspace{-0.05in} \begin{array}{cccc}  \\
 \end{array} \left(\delta_{k_z} \ \triangle z_c\right)^{\mathcal{N}_o}  \hspace{-0.05in} \begin{array}{cccc}  \\
 \end{array}\right],
 \end{eqnarray}

\vspace{-0.1in}

\begin{eqnarray}\label{deltakzexpr}
  \rightarrow  \delta_{k_z} \sim \frac{1}{\triangle z_c }   \left( \hspace{-0.05in} \begin{array}{cccc}  \\
 \end{array} \frac{1}{\gamma_{\hbox{\tiny TI}}} \hspace{-0.05in} \begin{array}{cccc}  \\
 \end{array} \right)^{-\frac{1}{\mathcal{N}_o}} \sim  \frac{1}{ \mathcal{N}_z \lambda }   \left( \hspace{-0.05in} \begin{array}{cccc}  \\
 \end{array} \frac{1}{\gamma_{\hbox{\tiny TI}}} \hspace{-0.05in} \begin{array}{cccc}  \\
 \end{array} \right)^{-\frac{1}{\mathcal{N}_o}},
\end{eqnarray}

The number of spectral reference planes $\mathcal{N}_r$ is given as

\vspace{-0.1in}

\begin{eqnarray}\label{Nr2}
 \mathcal{N}_r  =  \frac{\triangle k_{z,c}}{\delta_{k_z}}  \sim  2 \pi
  \alpha  \left( \hspace{-0.05in} \begin{array}{cccc}  \\
 \end{array} \frac{1}{\gamma_{\hbox{\tiny TI}}} \hspace{-0.05in} \begin{array}{cccc}  \\
 \end{array} \right)^{\frac{1}{ \mathcal{N}_o}}  \mathcal{N}_z,
\end{eqnarray}

The number of FFT operations is given as

\vspace{-0.1in}

\begin{eqnarray}\label{NFFT1}
 \mathcal{N}_{\hbox{\tiny FFT}}     \sim  2 \pi
  \alpha  \left( \hspace{-0.05in} \begin{array}{cccc}  \\
 \end{array} \frac{1}{\gamma_{\hbox{\tiny TI}}} \hspace{-0.05in} \begin{array}{cccc}  \\
 \end{array} \right)^{\frac{1}{ \mathcal{N}_o}}  \mathcal{N}_o
 \mathcal{N}_z.
\end{eqnarray}

It is obvious that $ \mathcal{N}_r$ in (\ref{Nr2}) and
$\mathcal{N}_{\hbox{\tiny FFT}}$ in (\ref{NFFT1}) are the same as
those given in   (\ref{Nr}) and (\ref{NFFT}), which also means that
the  spatial and spectral TI-FFTs have the same optimized
computational complexity.

\begin{figure}[t]\centering
\includegraphics[width=5.5 in, height= 4.3  in]{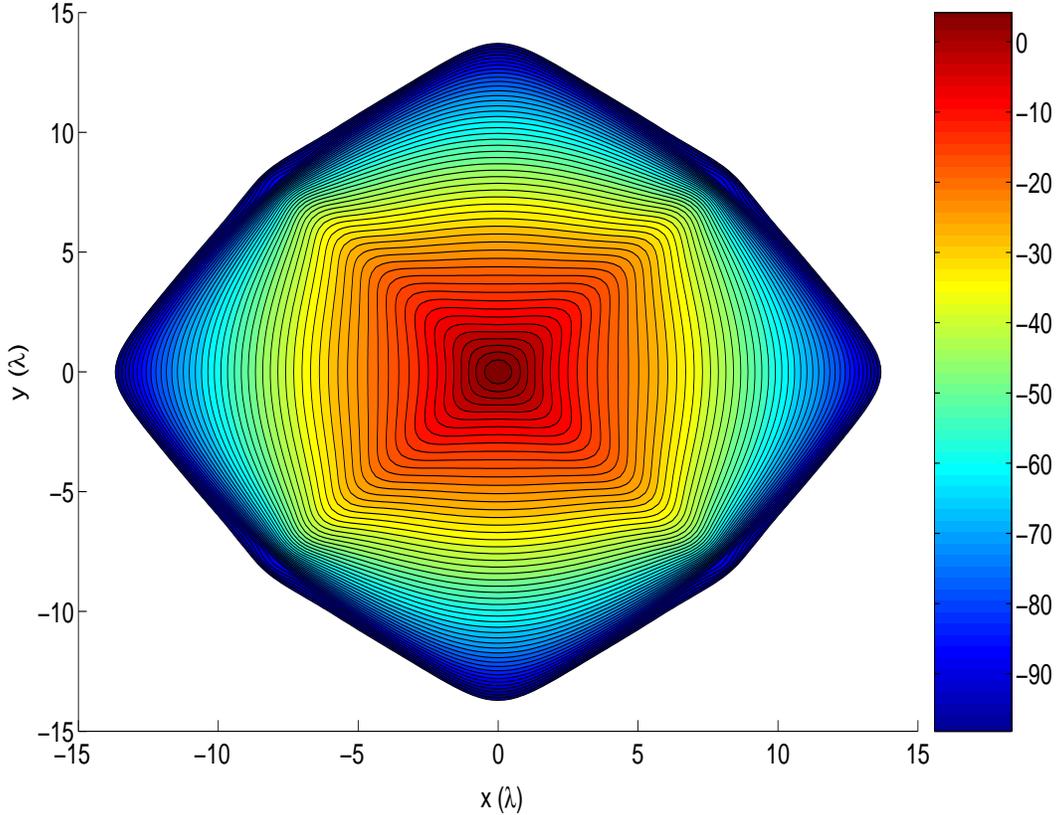}
\caption{ The  x-component (magnitude) of the  scattered  output
field.} \label{fig:Ex}
\end{figure}

\section{Computational Results}\label{results}

To show the efficiency of the planar TI-FFT algorithm, the direct
integration   of the radiation integral in (\ref{radiation}) has
been used to make comparison with the planar TI-FFT algorithm. The
numerical example used for such purpose is a 110 GHz Fundamental
Gaussian Beam (FGB) scattered by a PEC  quasi-planar surface with a
sin  wave perturbation. The 110 GHz FGB has a wavelength of $\lambda
\sim$ 2.7 mm.

 \begin{figure}[t]\centering
\includegraphics[width=5.8in, height= 4.5  in]{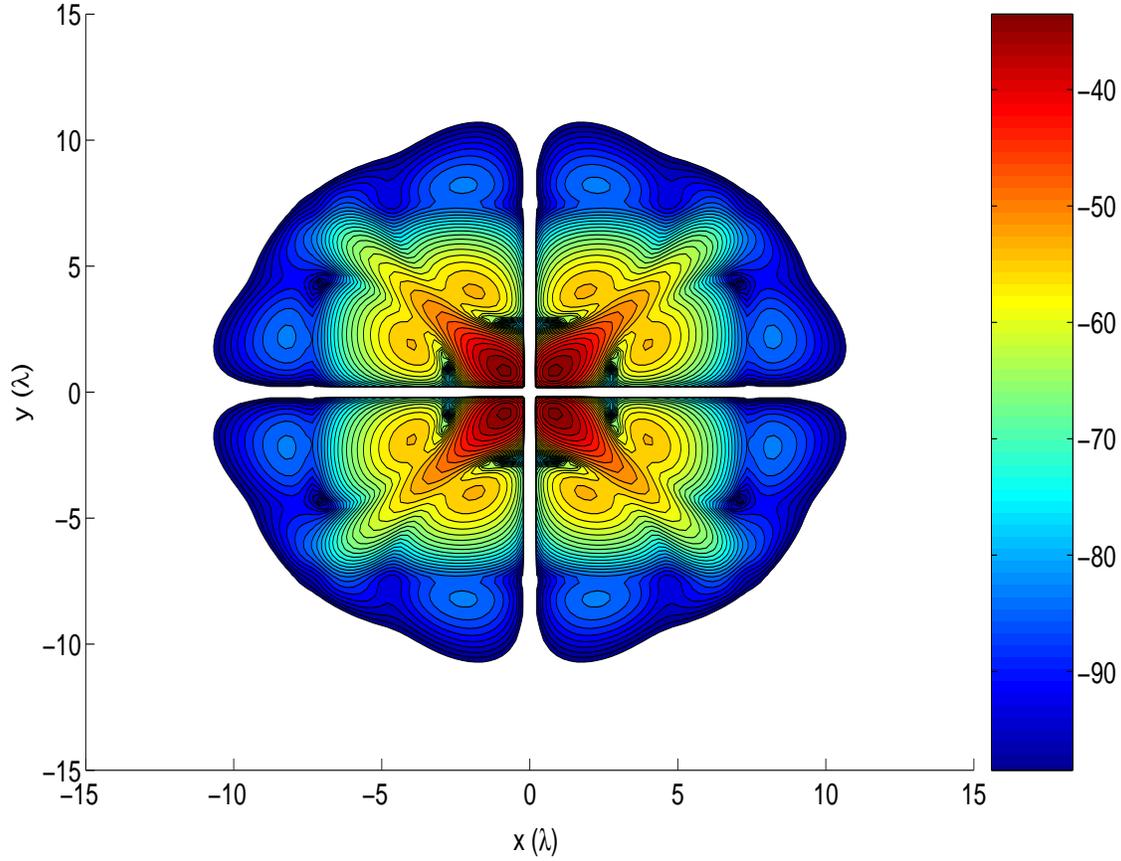}
\caption{ The  y-component (magnitude) of the  scattered  output
field.} \label{fig:Ey}
\end{figure}

\subsection{The numerical results}\label{Gaussian}

 The incident 110 GHz FGB propagates at $-\hat{\bf z}$ direction   and  has a beam
waist radius of $ w =  1 \ \hbox{cm}$.  The   quasi-planar PEC
surface with a sine wave perturbation  is described as

\vspace{-0.1in}

\begin{eqnarray}\label{quasi-planar}
 z(x,y)    =  -2.5 \lambda +  0.5 \lambda \cos\left(2 \pi
\frac{x}{15 \lambda}\right) \cos\left(2 \pi \frac{y}{15
\lambda}\right).
\end{eqnarray}

In the numerical implementation of the planar TI-FFT algorithm, the
computational accuracy $\gamma_{\hbox{\tiny TI}} = 0.0001$ ($-80$
dB) has been used and   the following optimized quantities are
obtained from (\ref{No1})-(\ref{complexityopt}),

\vspace{-0.1in}

\begin{eqnarray}\label{exp1}
\mathcal{N}_o^{\hbox{\tiny opt}}   \sim   9, \hspace{0.2in}
\delta_z^{\hbox{\tiny
  opt}}  \sim  0.6 \lambda, \hspace{0.2in}  \mathcal{N}_r^{\hbox{\tiny opt}}  \sim   \frac{1}{0.6} \sim
  2, \hspace{0.2in}  N_{\hbox{\tiny FFT}}^{\hbox{\tiny opt}}   \sim  18,   \hspace{0.1in}  \hbox{CPU}^{\hbox{\tiny opt}}
 \sim  18 \mathcal{O}  \left[\hspace{-0.05in} \begin{array}{cccc}  \\
 \end{array} N \log_2 N  \hspace{-0.05in} \begin{array}{cccc}  \\
 \end{array}\right].
\end{eqnarray}

\hspace{-0.22in}where the quasi-planar surface described in
(\ref{quasi-planar}) has a characteristic surface variation
$\triangle z_c \sim 1 \lambda$.

\begin{figure}[t]\centering
\includegraphics[width=5.8in, height= 4.5  in]{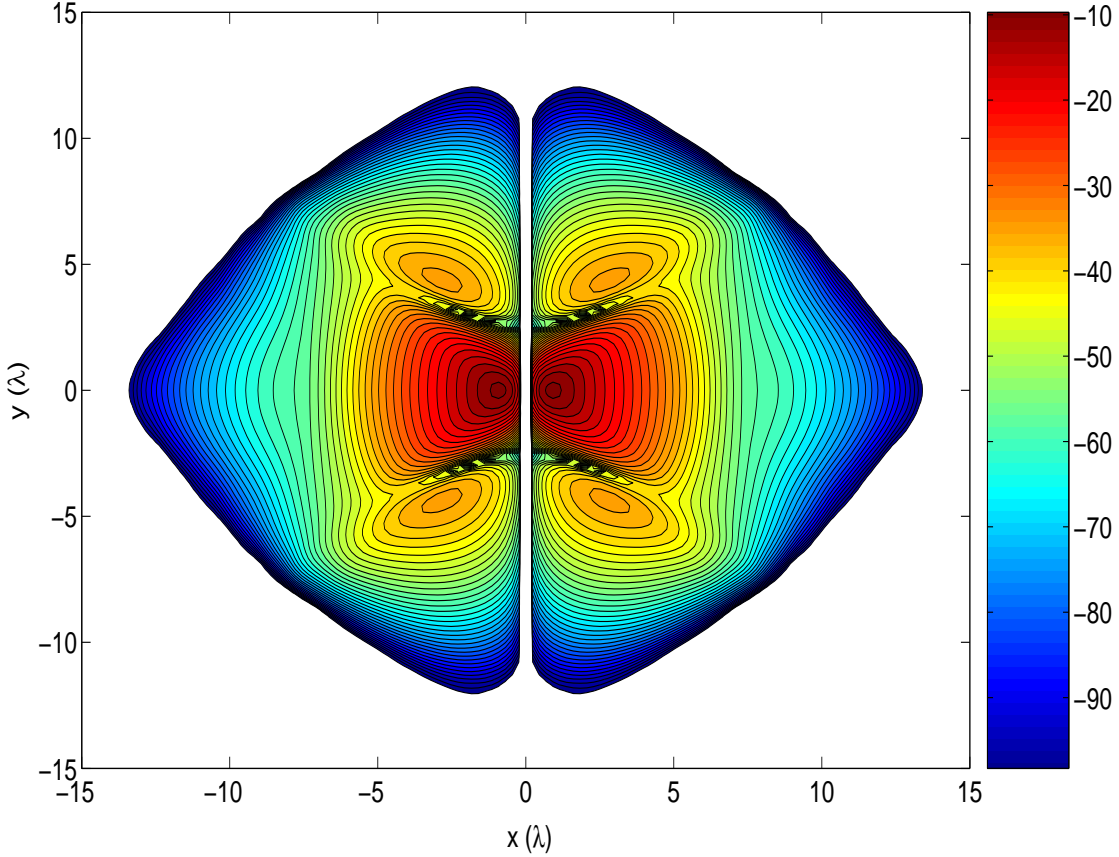}
\caption{ The  z-component (magnitude) of the  scattered  output
field.} \label{fig:Ez}
\end{figure}

\begin{figure}[t]
\hspace{-0.4in}\includegraphics[width= 5.5 in, height= 4.1
in]{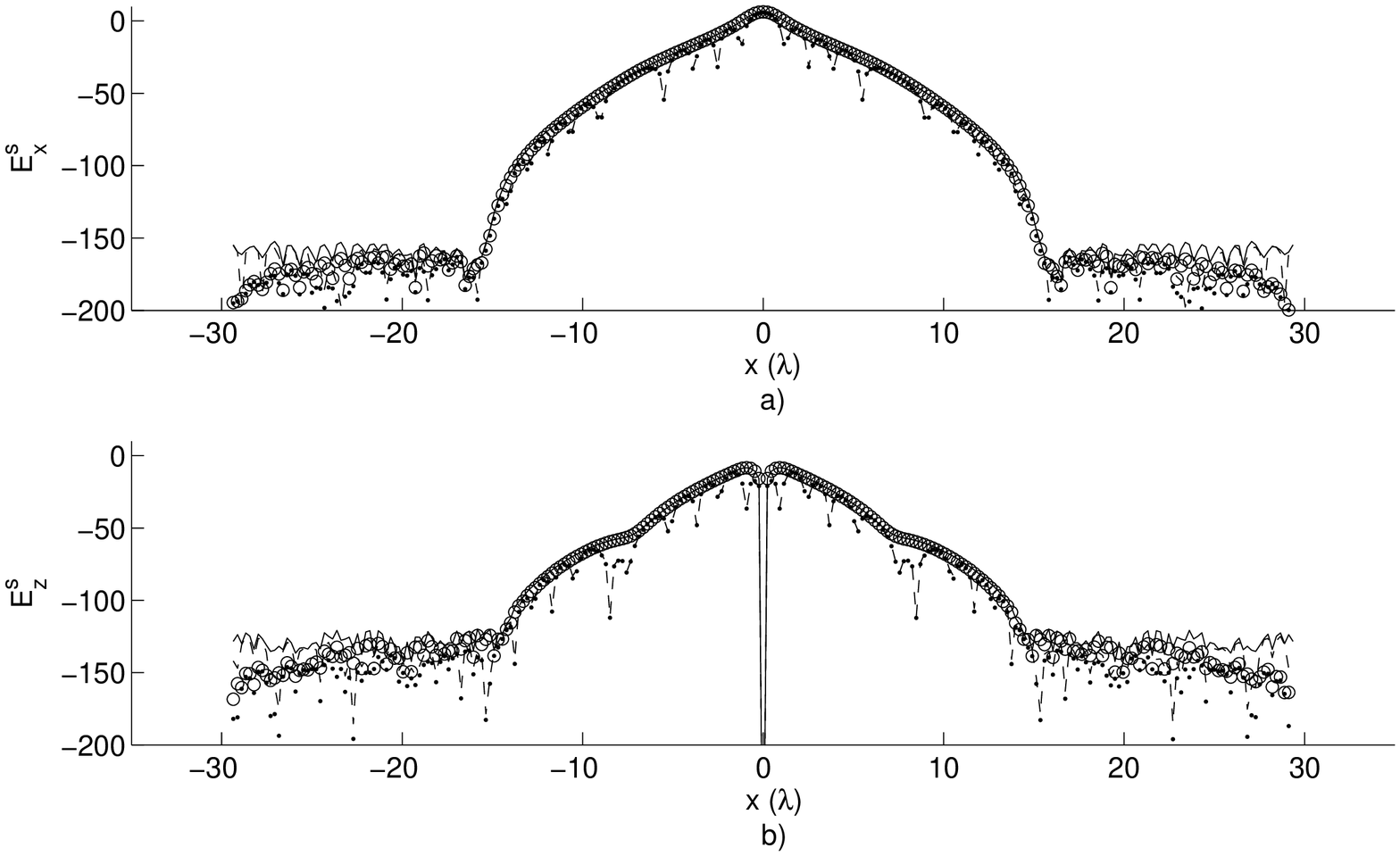} \caption{ The comparison of the scattered output
field  on plane $z=0$, across the maximum value point of
$|\hbox{E}_x^s|$ and at
  $\hat{\bf x}$ direction:  a) is for $\hbox{E}_x^s$; and b) is for $\hbox{E}_z^s$; solid lines (TI-FFT) and circles (direct
integration method) are magnitudes; dashed lines (TI-FFT) and dots
(direct integration method) are real parts.} \label{fig:comp}
\end{figure}

The scattered output field ${\bf E}^s$ are evaluated  on plane $z=0$
(where the incident 110 GHz FGB starts to propagate). Fig.
\ref{fig:Ex}, Fig. \ref{fig:Ey} and Fig. \ref{fig:Ez} show the
magnitude patterns of x-, y-, and z-components of the scattered
output field ${\bf E}^s$. The comparison of the result obtained from
the planar TI-FFT algorithm and that from the direct integration
method is given in Fig. \ref{fig:comp}, for both the magnitudes  and
the real parts, which shows that the planar TI-FFT algorithm has the
desired $-80$ dB computational accuracy.

\subsection{ The CPU time and the
accuracy}\label{CPU}

 The  CPU time t$_{ \hbox{\tiny   TI}}$ for the   planar TI-FFT algorithm  and t$_{ \hbox{\tiny  DI}}$ for the
direct integration method have been summarized in Table
\ref{Performance}, together with the coupling coefficient defined
 as

\vspace{-0.1in}

\begin{equation}
  \mathcal{C}_\tau \equiv   \left|  \frac{\int\!\!\int    {   \hbox{ E}_{\hbox{\tiny TI},\tau}^s
 [\hbox{E}_{\hbox{\tiny DI},\tau}^s]^\ast \ dx dy }}{\sqrt{\int\!\!\int    { |\hbox{E}_{\hbox{\tiny TI},\tau}^s|^2} \ dx dy }
\sqrt{\int\!\!\int   {  |\hbox{E}_{\hbox{\tiny DI},\tau}^s|^2} \ dx
dy }}\right|_{z=0,}
\end{equation}

\begin{table}[b]  \caption{CPU time
($\hbox{t}_{\hbox{\tiny TI}}$, t$_{\hbox{\tiny DI}}$) and coupling
coefficient $ \mathcal{C}_\tau $ } \label{Performance} \centering
\begin{tabular}{cccccccccc}
 \hspace{0.15in} $\mathcal{N}_{x,y}$ \hspace{0.15in} &  \hspace{0.15in} $\delta$($\lambda$)
\hspace{0.15in}& \hspace{0.15in} t$_{\hbox{\tiny TI}} (\hbox{sec.})
$   \hspace{0.15in}&  \hspace{0.15in} t$_{\hbox{\tiny DI}}
(\hbox{sec.}) $ \hspace{0.15in}& \hspace{0.15in} t$_{\hbox{\tiny
DI}}/\hbox{t}_{\hbox{\tiny TI}} $
  \hspace{0.15in} & \hspace{0.15in} $ \mathcal{C}_x $(\%)  \hspace{0.15in} &
  \hspace{0.15in}  $ \mathcal{C}_y $(\%)  \hspace{0.15in} & \hspace{0.15in} $ \mathcal{C}_z $ (\%) \\
\hline
$ 128  $ & 0.46  &  2   &  81    &  41   & 99.98  & 99.92  & 93.97  \\
$ 256  $ &   0.23 & 10   &  1289  &  129  &  99.99 & 99.99  & 99.99   \\
$ 512 $ &  0.12  &  44   &   20616 & 469   & 99.99 & 99.99  &  99.99 \\
$ 1024 $ &  0.06  &  194 &  329853 & 1700 & 99.99 &  99.99 & 99.99\\
   \hline   \\
\end{tabular}
\end{table}

\hspace{-0.22in}where,   $ \hbox{ E}_{\hbox{\tiny TI},\tau}^s$ and $
\hbox{ E}_{\hbox{\tiny DI},\tau}^s$ ($\tau=x, y, z$)  denote the
scattered output field components obtained from the planar TI-FFT
algorithm and the direct integration method respectively. From
T{\small ABLE} \ref{Performance}, it can be seen that, even though
at a large sampling spacing  $\delta   = 0.46 \lambda$ ($
\mathcal{N}_x = \mathcal{N}_y \sim
 128$), the coupling coefficients  are still well above
$90.00\%$.  At this sampling rate, the direct integration method
using Simpson's 1/3 rule is not accurate enough
\cite{Shaolin_conf,Rong,Perkins}. Also note that the coupling
coefficients $\mathcal{C}_\tau$ ($\tau=x,y,z$) reach their maximum
values of $99.99\%$ at $\mathcal{N}_x  = \mathcal{N}_y \sim 256$
($\delta_x = \delta_y = 0.23 \lambda$), after which the accuracies
remain constant and thus the Nyquist rate can be estimated roughly
as $\mathcal{N}_{\hbox{\tiny Nyquist}} \sim 256$. The reason for
this phenomenon is, that after the sampling rate increases above the
Nyquist rate, further increasing the sampling rate will not give
more information or computational accuracy.

\begin{figure}[t]
\includegraphics[width= 5.8 in, height= 4.0 in]{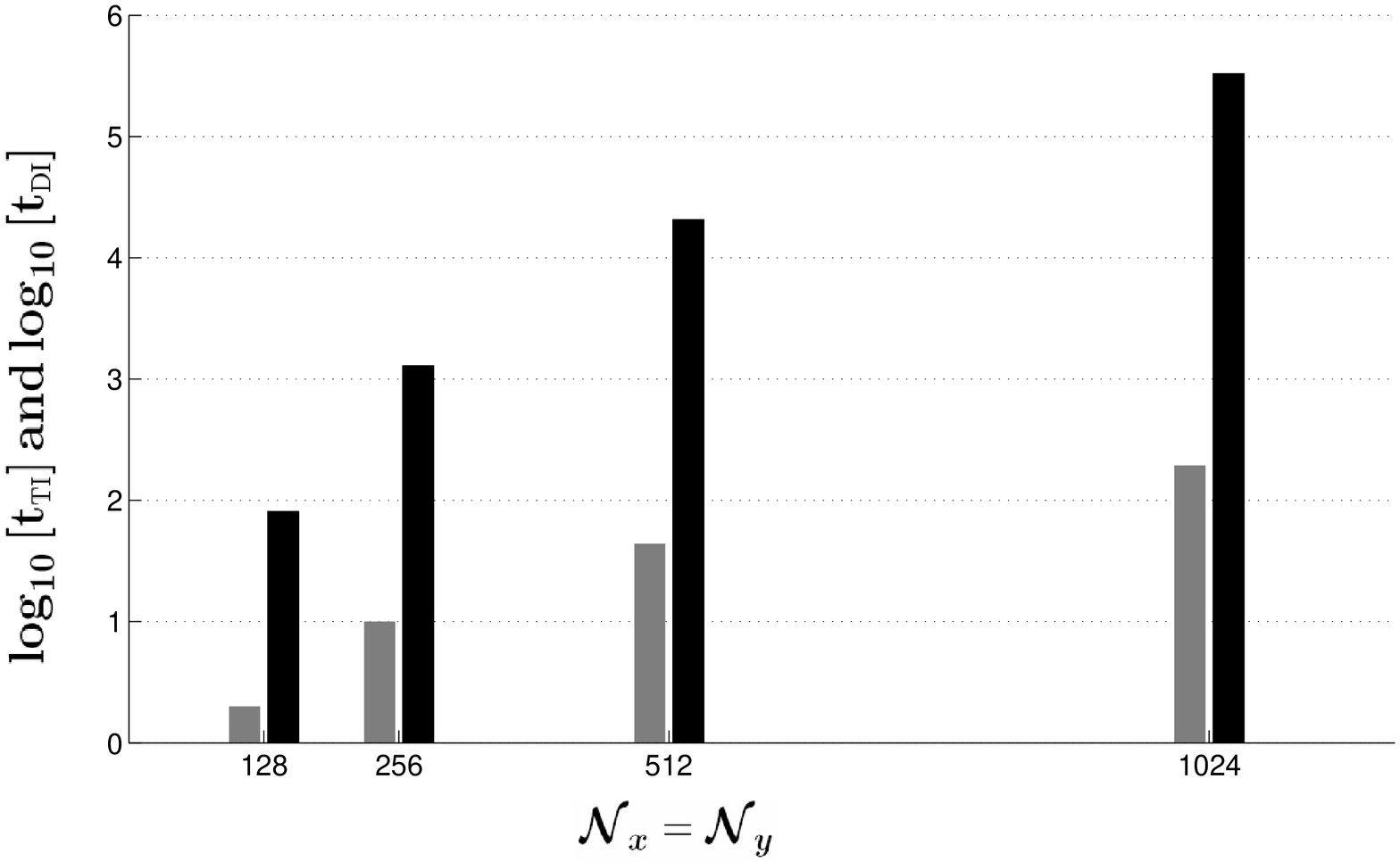}
\caption{ The   CPU time  (t$_{ \hbox{\tiny  TI}}$, t$_{ \hbox{\tiny
DI}}$) comparison: bars in gray color are for the planar TI-FFT
algorithm and bars in black color are  for the direct integration
method. Note that the CPU time is in logarithmic  scale (10-base).}
 \label{fig:cpu}
\end{figure}

\begin{figure}[t]
\includegraphics[width= 5.8 in, height= 4.0 in]{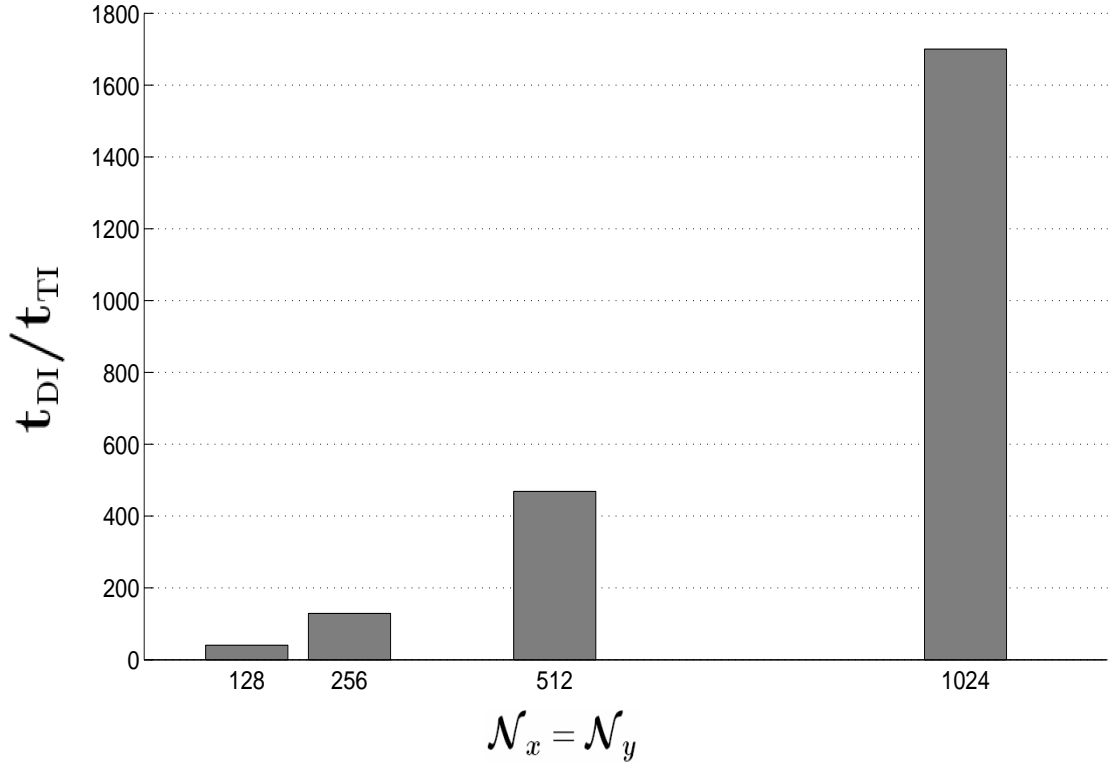}
\caption{ The efficiency of  the planar TI-FFT algorithm: the ratio
of t$_{ \hbox{\tiny DI}}/$t$_{ \hbox{\tiny TI}}$ for $\mathcal{N}_x
= \mathcal{N}_y = (128, 256, 512, 1024)$. } \label{fig:cpucmp}
\end{figure}

 The CPU time for the planar TI-FFT algorithm t$_{\hbox{\tiny TI}}$ and for
the direct integration method t$_{\hbox{\tiny DI}}$ are shown in
Fig. \ref{fig:cpu}. The ratio t$_{\hbox{\tiny DI}}/$t$_{\hbox{\tiny
TI}}$ is shown in Fig. \ref{fig:cpucmp}.

All work was done in Matlab 7.0.1, on a 1.66 GHz PC (Intel Core
Duo), with 512 MB Memory.

\section{Discussion: Problems and Possible Solutions} \label{sec:discussion}

Although the planar TI-FFT algorithm has so many advantages given
above, some problems do exist in the practical applications.

\begin{figure}[h] \centering
\includegraphics[width= 4.  in, height= 4.  in]{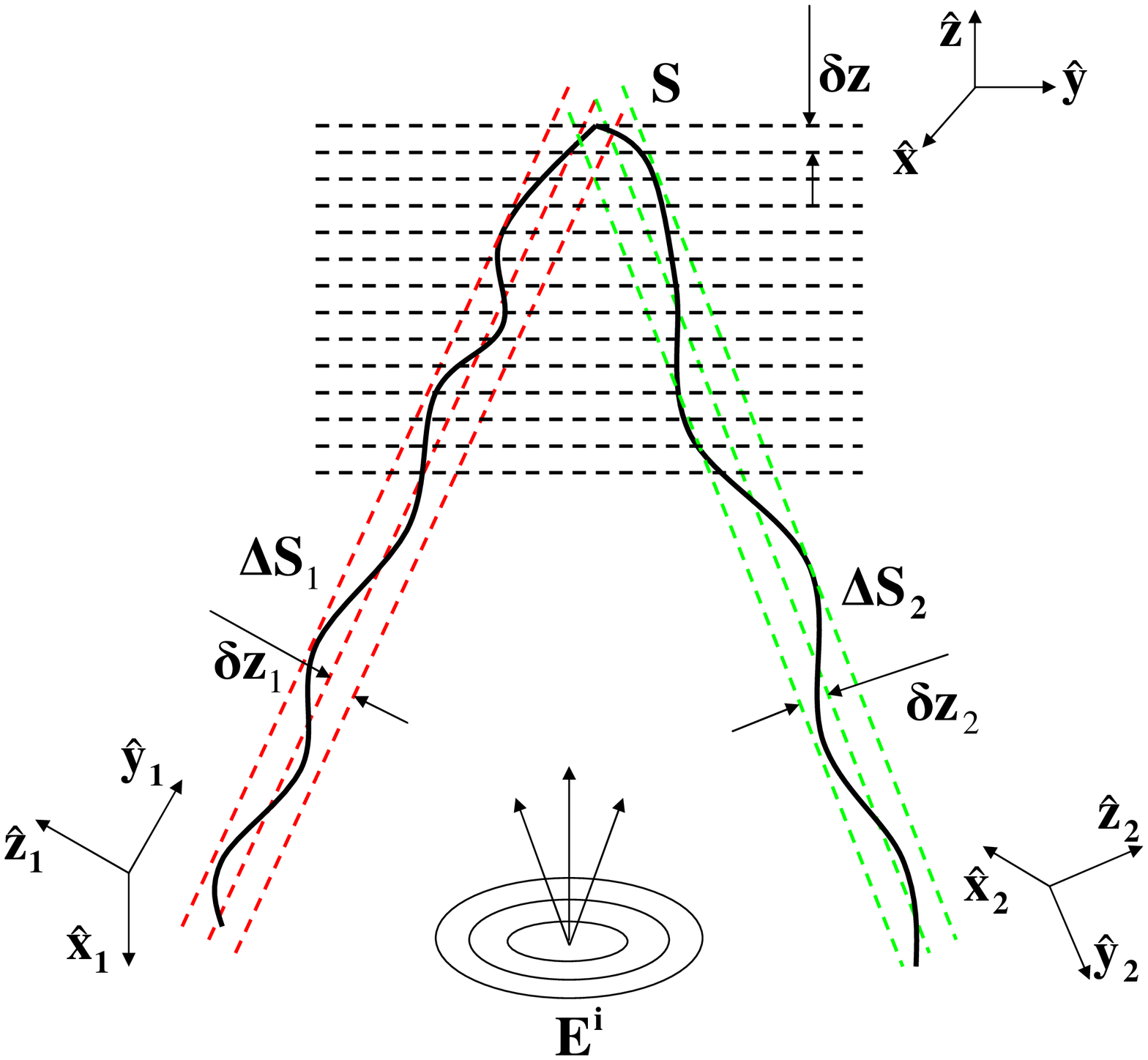}
\caption{An example of complicate surface $S$ that can be divided
into two quasi-planar surface patches $\triangle S_1$ and $\triangle
S_2$. The computations of each surface patch is done in its
corresponding coordinate system whose z-coordinate is perpendicular
to the slicing spatial reference planes.}
 \label{fig:complicate}
  \end{figure}

\subsubsection{Complicate geometry}
\label{geometry}

As an example, consider surface $S$ shown in Fig.
\ref{fig:complicate}, where the surface itself is not a quasi-planar
surface and the direct implementation of the planar TI-FFT algorithm
requires a large number of FFT operations, which can be seen from
the spatial reference planes with a spatial slicing spacing
$\delta_z$. The problem can be solved by dividing  surface $S$ into
two  surface patches $\triangle S_1$ and $\triangle S_2$, which can
be considered as quasi-planar surfaces and the planar TI-FFT can be
used on them independently, with coordinate systems selected based
on the spatial reference planes.  At the extreme limit where surface
patches $\triangle S_1$ and $\triangle S_2$ are planes, the number
of FFT operations reduces to $\mathcal{N}_{\hbox{\tiny FFT}} = 2$.

\begin{figure}[t] \centering

\includegraphics[width= 3.8  in, height= 1.8 in]{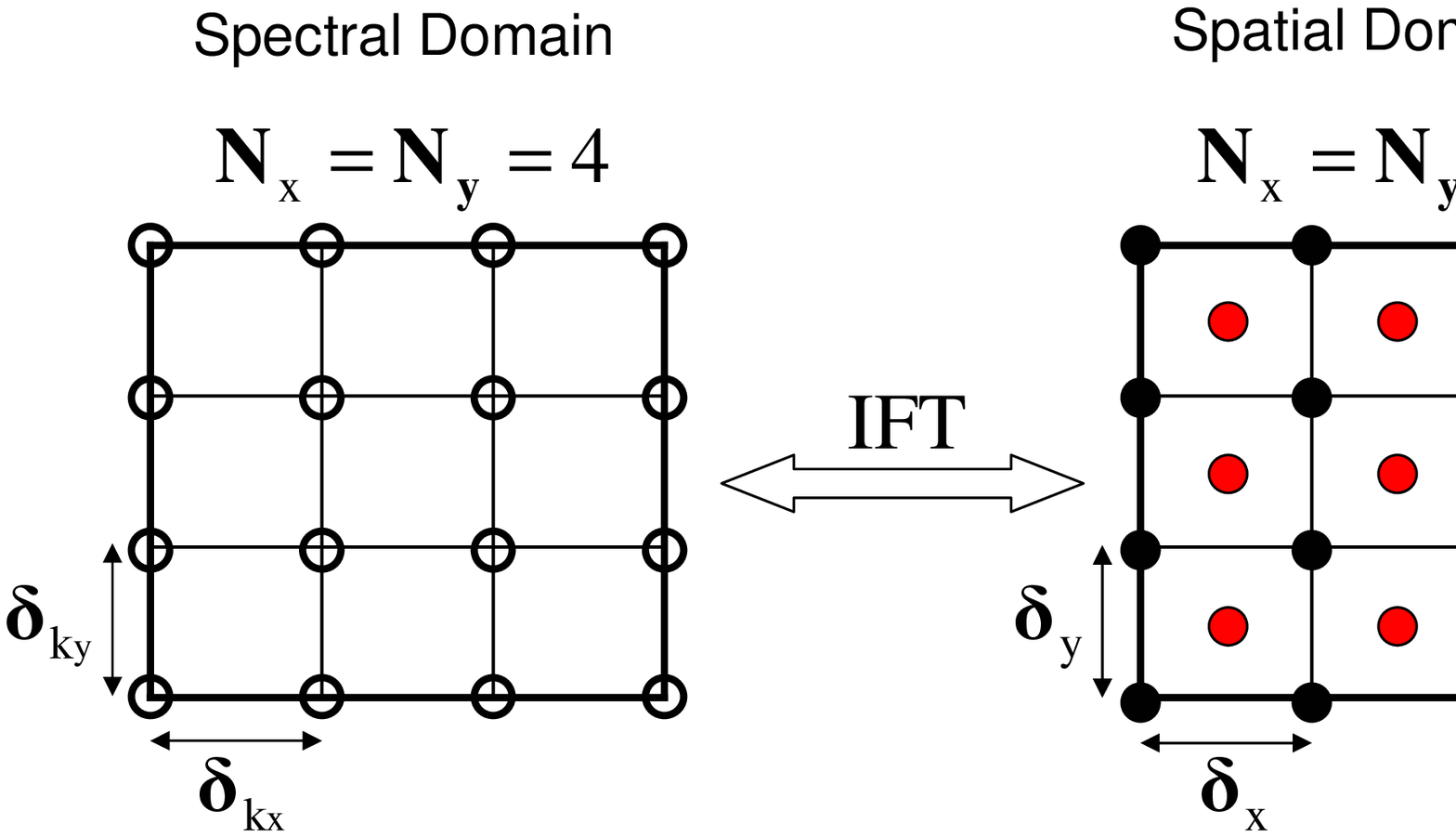}
\caption{ The problem of computation of electromagnetic field on the
observation points  that are not on the computational grid ($4
\times 4$), which are denoted as red filled circles in the spatial
domain (assume that they are evenly distributed). ($\delta_{k_x}$,
$\delta_{k_y}$) are grid spacings in the spectral domain.
($\delta_x$, $\delta_y$) are grid spacings in the space domain. }
 \label{fig:4grid}

\vspace{0.2in}
 \includegraphics[width=4.7 in, height= 3.4 in]{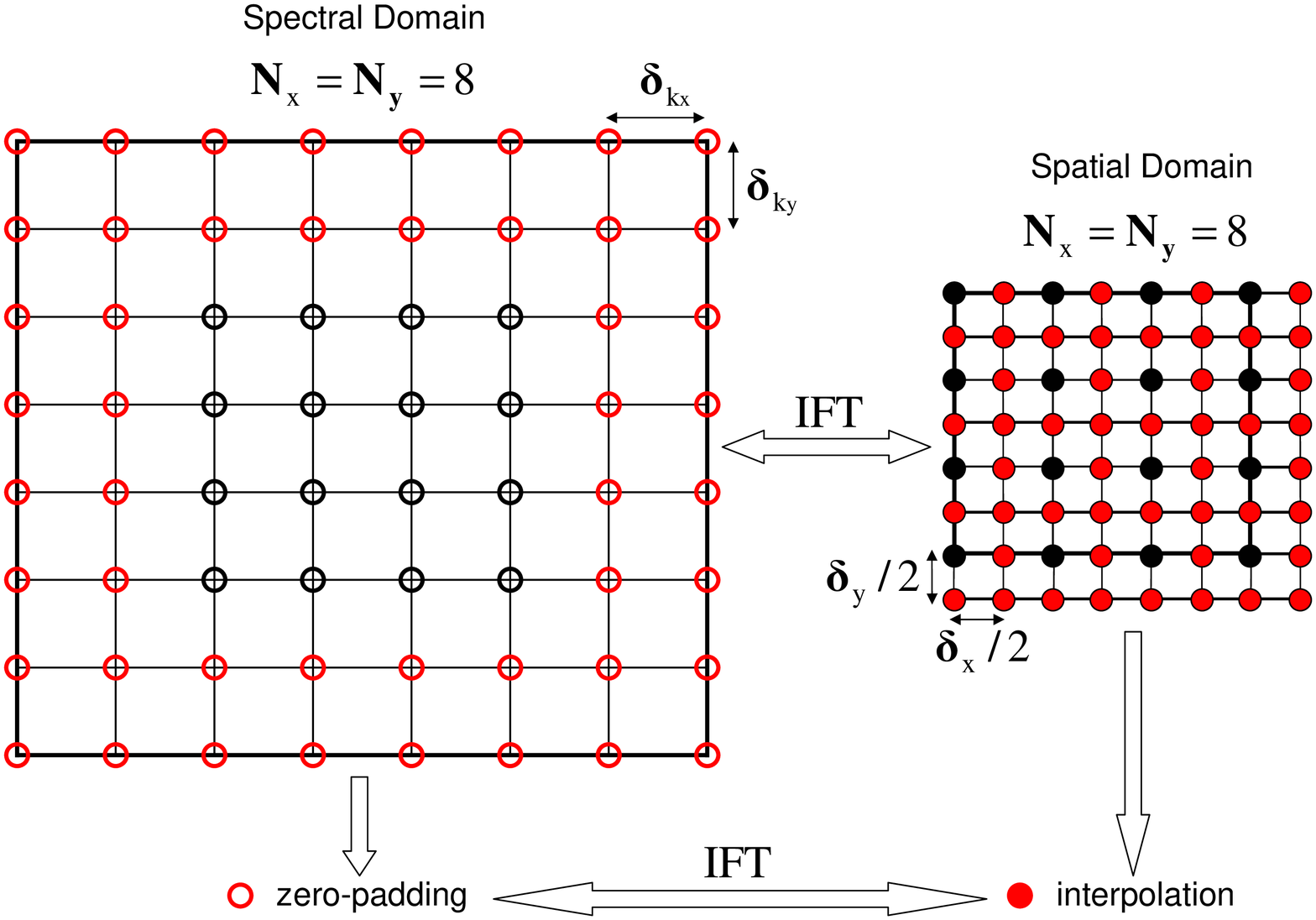}
\caption{ The zero-padding   in the spectral domain ($4 \times 4
\rightarrow 8 \times 8$) corresponding to the interpolation in the
spatial domain ($4 \times 4 \rightarrow 8 \times 8$).
($\delta_{k_x}$, $\delta_{k_y}$) are still the same after
zero-padding. But   grid spacings in the spatial domain become
($\delta_x/2$, $\delta_y/2$) after interpolation.}
 \label{fig:8grid}

  \end{figure}

\subsubsection{Observation points not on the computational grid}
\label{grid}

It is well-known that the FFT requires an even grid spacing (but
$\delta_x$ and $\delta_y$ need not to be equal), which raises the
question of how to calculate the electric field at points that are
not exactly on the computational grid, e.g., the red filled circles
in Fig. \ref{fig:4grid}. One solution for this problem is to
zero-pad the computational grid in the spectral domain, which
corresponds to the interpolation of the computational grid in the
spatial domain, as shown in Fig. \ref{fig:8grid}. In the above
example, it has been assumed that  the observation points are evenly
distributed and the interpolation results are exact provided that
the sampling rate is above the Nyquist rate \cite{Oppenheim}.  For
complicate observation point configurations (e.g., unevenly
distributed points), the approximate techniques like the Gauss's
forward/backward interpolations can be used.

\begin{figure}[h] \centering
\includegraphics[width= 5.0 in, height= 4.3 in]{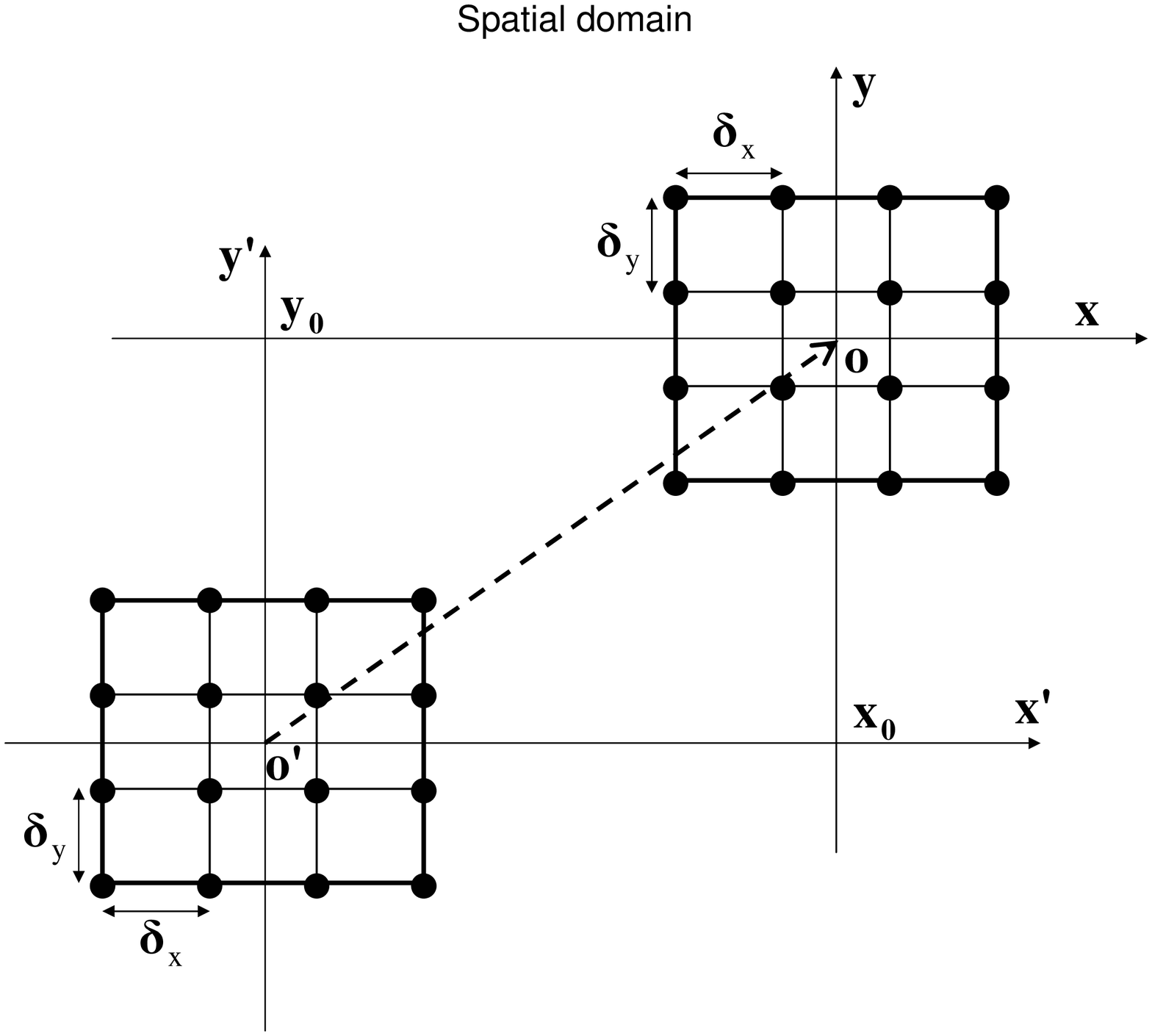}
\caption{The translation of the source coordinate system $o'$(0,0)
to the observation coordinate system $o$($x_0$, $y_0$) in the
spatial domain. Both the source  and observation coordinate systems
should have the same grid spacings ($\delta_x$, $\delta_y$).}
 \label{fig:translation}

  \end{figure}

\subsubsection{The translation in spatial domain}
\label{translation}

In the real situation, the source field surface and the observation
surface are separate far away from each other (see Fig.
\ref{fig:translation}). It is not practical nor necessary to use a
large computational grid that covers both the source field surface
and the observation surface. This kind of problem can be solved by
using two computational grids, one for the source field surface and
the other for the observation surface, with the same grid spacings
($\delta_x$, $\delta_y$). Then the translation of the observation
coordinate system in the spatial domain, which is denoted as ($x_0$,
$y_0$), corresponds to the phase shift in the spectral domain.
Suppose the electric field in the source coordinate system is
expressed as ${\bf E}(x' - x_0, y'- y_0)$, according to the property
of the Fourier transform \cite{Oppenheim}, the electric field ${\bf
E}(x, y) $ in the observation coordinate system is given as

\vspace{-0.2in}

\begin{eqnarray}\label{translation}
  \hspace{-0.9in} {\bf E}(x, y)     = \hbox{\large IFT}_{\hbox{\tiny 2D}} \left[
\hspace{-0.05in}
\begin{array}{cc} \\ \\ \end{array}  e^{-j k_x x_0} e^{-j k_y y_0} \hspace{-0.05in}
\begin{array}{cc} \\ \end{array}   \hbox{\large FT}_{\hbox{\tiny 2D}} \left[\hspace{-0.05in}
\begin{array}{cc} \\ \end{array}  {\bf E}(x' - x_0, y'- y_0)   \hspace{-0.05in}
\begin{array}{cc} \\  \\ \end{array}
\right]  \right].
\end{eqnarray}

\subsubsection{Computational redundancy}
\label{redundancy}

In the numerical implementation of the planar TI-FFT algorithm, the
spatial domain or the spectral domain are divided into many small
subdomains where the FFT can be used to interpolate the
electromagnetic field (see Fig. \ref{fig:scheme} and Fig.
\ref{fig:spectral}). However, the FFT operation is done on the whole
spatial or spectral domain even though  the interpolation is only
necessary on the relatively small subdomain, which causes the
computational redundancy in the planar TI-FFT algorithm.
Fortunately,  the computational redundancy  is small for a
quasi-planar surface  and a narrow-band beam.

\section{Conclusion}\label{sec:conclusion}

In this article, the optimized planar TI-FFT algorithm for the
computation of electromagnetic wave propagation   has been
introduced for   the narrow-band beam and the quasi-planar geometry.
Two types of TI-FFT algorithm are available, i.e.,  the {spatial}
TI-FFT  and   the {spectral} TI-FFT. The former is  for computation
of electromagnetic wave on the quasi-planar surface and the latter
is for computation of the 2D Fourier spectrum of the electromagnetic
wave.    The optimized order of Taylor series used in the planar
TI-FFT algorithm is found to be closely related to the algorithm's
computational accuracy $\gamma_{\hbox{\tiny TI}}$, which is given as
$\mathcal{N}_o^{\hbox{\tiny opt}} \sim - \ln \gamma_{\hbox{\tiny
TI}}$ and  the optimized spatial slicing spacing between two
adjacent spatial reference planes  only depends on the
characteristic wavelength  $\lambda_c$   of the electromagnetic
wave, which is $\delta_z^{\hbox{\tiny opt}} \sim \frac{1}{17}
\lambda_c$. The optimized computational complexity is given as $
\mathcal{N}_r^{\hbox{\tiny opt}} \mathcal{N}_o^{\hbox{\tiny opt}}
 \mathcal{O} \left(N \log_2 N \right)$  for an $N = \mathcal{N}_x
\times \mathcal{N}_y$ computational grid. The planar TI-FFT
algorithm allows a low sampling rate (large sampling spacing)
required by the sampling theorem. Also, the algorithm doesn't have
the problem of singularity. The planar TI-FFT algorithm has
applications in near-field and far-field computations, beam-shaping
mirror system designs, diffraction and scattering phenomena,
millimeter wave propagation, and microwave imaging in the half-space
scenario.

\section*{Acknowledgment}
 This work was supported by the U.S. Dept. of Energy under the
contract DE-FG02-85ER52122.

\end{document}